\documentclass[twoside]{IEEEtran}
\ifCLASSINFOpdf
   \usepackage{graphicx}
    \usepackage[outdir=./]{epstopdf}
\else
   \usepackage[dvips]{graphicx}
   \graphicspath{{/Users/swanandkadhe/Dropbox/LRCs/Local_MRD/Local_MRD_main/}}
  \DeclareGraphicsExtensions{.eps}
\fi
\usepackage{amsfonts, amssymb, amscd, amsthm, amsmath, xspace}
\usepackage{subfigure}

\usepackage{algorithm}
\usepackage{algorithmic}
\usepackage{xfrac,xcolor}
\usepackage{flushend}

\newtheorem{theorem}{Theorem}
\newtheorem{lemma}{Lemma}

\newtheorem{corollary}{Corollary}
\newtheorem{remark}{Remark}
\newtheorem{construction}{Construction}

\newtheorem{example}{Example}
\newtheorem{definition}{Definition}
\newtheorem{proposition}{Proposition}

\newcommand{\etal}{{\it et al.}}
\newcommand{\ie}{{\it i.e.}}
\newcommand{\eg}{{\it e.g.}}

\newcommand{\twomatrix}[2]{\begin{bmatrix} #1 \\ #2\end{bmatrix}}

\newcommand{\GF}[1]{\mathbb{F}_{#1}} 

\newcommand{\code}{\mathcal{C}} 
\newcommand{\codeproj}[1]{\mathcal{C}\mid_{#1}} 

\newcommand{\vect}[1]{\mathbf{#1}} 
\newcommand{\cw}{\mathbf{c}} 

\newcommand{\wt}[1]{\textsf{wt}\left(#1\right)} 
\newcommand{\wtc}[1]{\textsf{wt}_c\left(#1\right)} 
\newcommand{\dims}[1]{\textsf{dim}\left(#1\right)} 
\newcommand{\rnk}[1]{\textsf{rank}\left(#1\right)} 

\newcommand{\gam}[1]{\Gamma\left(#1\right)} 
\newcommand{\subspace}[1]{\left\langle #1 \right\rangle} 

\newcommand{\cwset}[1]{c_{#1}} 

\newcommand{\ceillr}[1]{\left\lceil#1\right\rceil} 

\newcommand{\xq}[2]{{#1}^{[#2]}} 

\newcommand{\setP}{P} 
\newcommand{\setPpartition}{\mathcal{P}} 

\newcommand{\GFm}[1]{\mathbb{F}_{{#1}^m}} 
\newcommand{\GFmn}[1]{\mathbb{F}_{#1}^{m\times n}} 
\newcommand{\GFext}[2]{\mathbb{F}_{{#1}^{#2}}} 

\newcommand{\basis}[1]{\mathcal{#1}} 

\newcommand{\al}[1]{\alpha_{#1}} 
\newcommand{\be}[1]{\beta_{#1}} 

\newcommand{\dr}[1]{d_R\left(#1\right)} 
\newcommand{\dm}[1]{d_{min}\left(#1\right)} 
\newcommand{\degq}[1]{\textsf{deg}_q\left(#1\right)} 
\newcommand{\ds}[1]{d_S\left(#1\right)} 

\newcommand{\m}{\vect{m}} 
\newcommand{\mij}[1]{m_{#1}} 
\newcommand{\eij}[1]{e_{#1}} 
\newcommand{\wij}{\omega_{ij}} 

\newcommand{\Pq}[1]{\mathcal{P}_q\left(#1\right)} 
\newcommand{\Gq}[2]{\mathcal{G}_q\left(#1,#2\right)} 
\newcommand{\mymatrix}[1]{\left[ #1 \right]} 
\newcommand{\project}[2]{{#1\mid}_{#2}} 
\newcommand{\rcef}[1]{\textsf{rcef}\left(#1\right)} 

\newcommand{\Err}{E'} 
\newcommand{\erij}[1]{e'_{#1}} 
\newcommand{\GFM}[1]{\mathbb{F}_{q}^{#1}} 

\begin{document}
%
\title{Codes with Locality in the Rank and Subspace Metrics}
%
%
%

\author{Swanand Kadhe, 
        Salim El Rouayheb, 
        Iwan Duursma,~
        and~Alex Sprintson
\thanks{This paper was presented in part at the 54th Annual Allerton Conference on Communication, Control, and Computing, Oct 2016.}
\thanks{Swanand Kadhe is with the Department of Electrical Engineering and Computer Sciences, University of California Berkeley, USA (e-mail: swanand.kadhe@berkeley.edu). Part of this work was done while he was with the ECE department at Texas A\&M University.}
\thanks{Salim El Rouayheb is with the Department of Electrical and Computer Engineering, Rutgers University, USA (e-mail: salim.elrouayheb@rutgers.edu.)}
\thanks{Iwan Duursma is with the Department of Mathmatics, University of Illinois Urbana-Champaign, USA (e-mail: duursma@illinois.edu.)}
\thanks{Alex Sprintson is with the Department of Electrical and Computer Engineering, Texas A\&M University, USA (e-mail: spalex@tamu.edu)}
\thanks{The work of I. Duursma was supported in part by the Simons Foundation under Grant 280107 and in part by NSF under Grant CCF-1619189. The work of A. Sprintson is supported in part by the National Science Foundation under Grants No 1718658 and 1642983.}
\thanks{Part of this work was done during the authors' stay at the Institut Henri Poincar\'e - Centre \'Emile Borel. The authors thank this institution for hospitality and support.}
}

\maketitle

\vspace{-2mm}

\begin{abstract}
We extend the notion of locality from the Hamming metric to the rank and subspace metrics. Our main contribution is to construct a class of array codes with locality constraints in the rank metric. Our motivation for constructing such codes stems from the need to design codes for efficient data recovery from  correlated and/or mixed (\ie, complete and partial) failures in distributed storage systems. Specifically, the proposed local rank-metric codes can recover locally from \emph{crisscross errors and erasures}, which affect a limited number of rows and/or columns of the storage array. We also derive a Singleton-like upper bound on the minimum rank distance of (linear) codes with \emph{rank-locality} constraints. Our proposed construction achieves this bound for a broad range of parameters. The construction builds upon Tamo and Barg's method for constructing locally repairable codes with optimal minimum Hamming distance. Finally, we construct a class of {constant-dimension subspace codes (also known as Grassmannian codes)} 
with locality constraints in the subspace metric. The key idea is to show that a {Grassmannian code} with locality can be easily constructed from a rank-metric code with locality by using the lifting method proposed by Silva \emph{et al}. We present an application of such codes for distributed storage systems, wherein nodes are connected over a network that can introduce errors and erasures.
\end{abstract}


%
\IEEEpeerreviewmaketitle

\section{Introduction}
\label{sec:intro}
Distributed storage systems have been traditionally  replicating data over multiple nodes to guarantee reliability against failures and protect the data from being lost~\cite{Rowstron:01, Ghemawat:03:GFS}. However, the enormous growth of data being stored or computed online has motivated practical systems to employ erasure codes for handling failures (\eg,~\cite{Huang:12, Muralidhar:14}). This has galvanized  significant interest in the past few years on novel erasure codes that efficiently handle node failures in distributed storage systems. One of the main families of codes that has received primary research attention is {\it locally repairable codes} (LRCs) -- that minimize locality, \ie, the number of nodes participating in the repair process (see, \eg,~\cite{Huang:07, Gopalan:12, Oggier:11, Sathiamoorthy:13, TamoB:14}). Almost all the work in the literature on LRCs has considered block codes under the Hamming metric.

In this work, we first focus our attention to codes with locality constraints in the rank metric. Let $\GF{q}$ be the finite field of size $q$. Codewords of a rank-metric code (also known as an array code) are $m\times n$ matrices over $\GF{q}$, where the rank distance between two matrices is the rank of their difference~\cite{Delsarte:78,Gabidulin:85,Roth:91}. 
We are interested in rank-metric codes with locality constraints. To quantify the requirement of locality under the rank metric, we introduce the notion of {\it rank-locality}. We say that the $i$-th column of an $m\times n$ array code has $(r,\delta)$ {\it rank-locality} if there exists a set $\gam{i}$ of $r+\delta-1$ columns containing $i$ such that the array code formed by deleting the columns outside $\gam{i}$ for each codeword has rank distance at least $\delta$. We say that an $m\times n$ array code has $(r,\delta)$ rank-locality if every column has $(r,\delta)$ rank-locality.

Our motivation of considering rank-locality is to design codes that can locally recover from {\it rank errors and erasures}. Rank-errors are the error patterns such that the rank of the error matrix is limited. For instance, consider an error pattern added to a codeword of a binary $4 \times 4$ array code as shown in Fig.~\ref{fig:crisscross}. Though this pattern  corrupts half the bits, its rank over the binary field is only one. 
\begin{figure}[!h] 
\begin{center}
$E \:\: = \:\:$
\begin{tabular}{|c|c|c|c|}
\hline
1 & 1 & 1 & 1\\
\hline
0 & 0 & 0 & 0\\
\hline
1 & 1 & 1 & 1\\ 
\hline
0 & 0 & 0 & 0\\
\hline
\end{tabular}
\end{center}
\caption{A rank-error pattern of rank one.}
\label{fig:crisscross}
\end{figure}
Note that it is not possible to correct such an error pattern using a code equipped with the Hamming metric. On the other hand, rank-metric codes are well known for their ability to effectively correct rank-errors~\cite{Roth:91,Blaum:00}. 

\begin{figure*}[!t]
\centering
\includegraphics[scale=0.5]{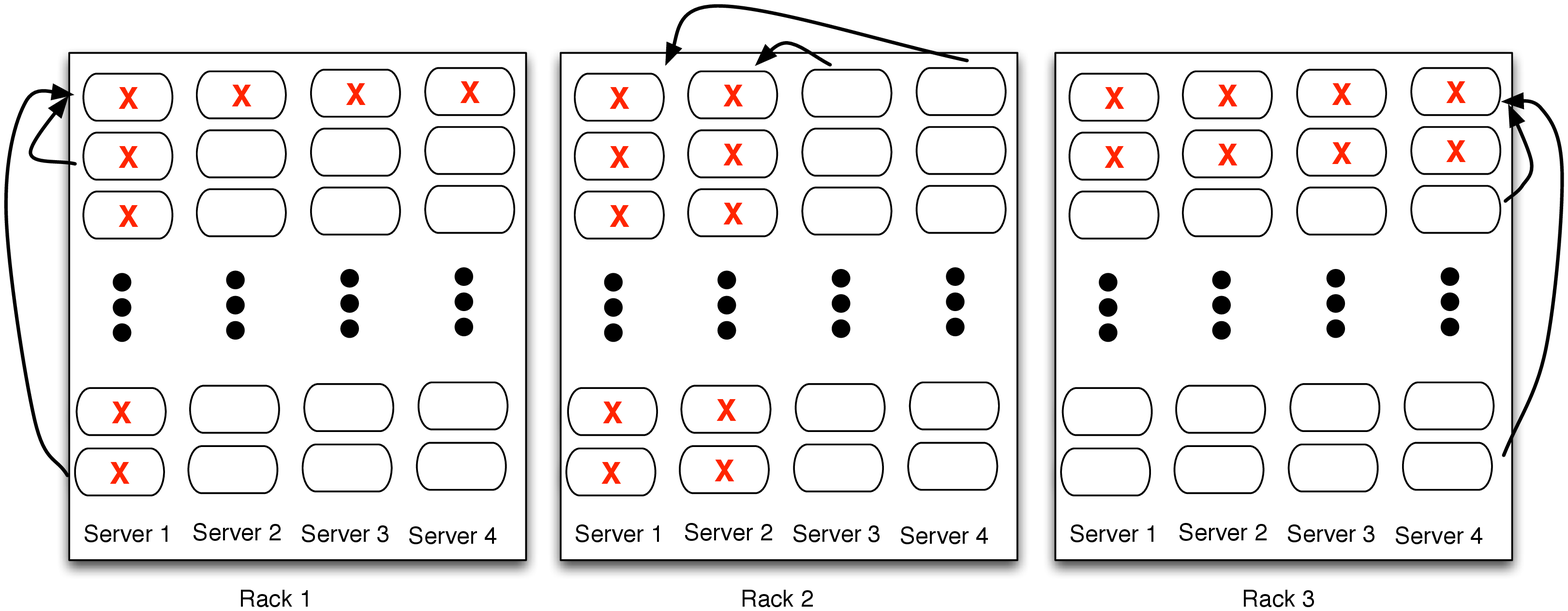}
\caption{Our motivation is to study codes for distributed storage systems that can locally recover from correlated and/or mixed failures, with particular focus on their subclass called crisscross failures. A crisscross failure pattern affects a limited number of rows and/or columns. For example, a few instances of crisscross failures affecting two rows and/or columns are depicted in the figure. We study rank-metric codes with local recoverability property as follows: any crisscross failure pattern that affects fewer than $\delta$ rows and/or columns of a rack can be locally recovered by accessing only the nodes in the same rack. 
}
\label{fig:9by9-array-erasures}
\end{figure*}

Errors and erasures that affect a limited number of rows and/or columns are usually referred to as {\it crisscross patterns}~\cite{Roth:91, Blaum:00}. 
(See Fig.~\ref{fig:9by9-array-erasures} for some examples of crisscross erasures.) Our goal is to investigate codes that can {\it locally} recover from crisscross erasures (and rank-errors). We note that crisscross errors (with no locality) have been studied previously in the literature~\cite{Roth:91, Blaum:00}, motivated by applications in memory chip arrays and multi-track magnetic tapes. Our renewed interest in these types of failures stems from the fact that they form a subclass of {\it correlated and mixed failures}, see, \eg,~\cite{Ford:Google:10, Gill:11}.

Recent research has shown that many distributed storage systems suffer from a large number of  correlated and mixed failures~\cite{Ford:Google:10, Gill:11, Nath:06, Bairavasundaram:08, Hafner:08, Balakrishnan:10}. For instance, a correlated failure of several nodes can occur due to, say, simultaneous upgrade of a group of servers, or a failure of a rack switch or a power supply shared by several nodes~\cite{Ford:Google:10,Gill:11,Nath:06}. Moreover, in distributed storage systems composed of solid state drives (SSDs), it is not uncommon to have a failed SSD along with a few corrupted {\it blocks} in the remaining SSDs, referred to as mixed failures~\cite{Balakrishnan:10,Blaum:13,Greenan:09}. Therefore, recent research on coding for distributed storage has also started focusing on correlated and/or mixed failure models, see~\eg,~\cite{Blaum:13,Shum:13,Rawat:14:Cooperative,Gopalan:14,Prakash:14,Gopalan:17:SODA,Plank:14,Blaum:16}.

Another potential application for codes with rank-locality is for correcting errors occurring in dynamic random-access memories (DRAMs). In particular, a typical DRAM chip contains several internal {\it banks}, each of which is logically organized into {\it rows} and {\it columns}. Each row/column address pair identifies a {\it word} composed of several bits. Recent studies show that DRAMs suffer from non-negligible percentage of bit errors, single-row errors, single-column errors, and single-bank errors~\cite{Sridharan:12, Sridharan:13, Sridharan:15}.  Using an array code across banks, with a local code for each bank can be helpful in correcting such error patterns.

In general, our goal is to design and analyze codes that can {\it locally recover} the crisscross erasure and error patterns, which affect a limited number of rows and columns, by accessing a small number of nodes. 
We show that a code with $(r,\delta)$ rank-locality can locally repair any crisscross erasure pattern that affects {fewer than} $\delta$ rows and columns by accessing only $r$ columns. We begin with a toy example to motivate the coding theoretic problem that we seek to solve.

\begin{example}
\label{ex:9by9-array}
Consider a toy example of a storage system, such as the one depicted in Fig.~\ref{fig:9by9-array-erasures}, consisting of three racks, each containing four servers. Each server is composed of several storage nodes which can either be solid state drives (SSDs) or hard disk drives (HDDs).\footnote{Many practical storage systems such as Facebook's `F4' storage system~\cite{Muralidhar:14} and all-flash storage arrays such as~\cite{Colgrove:15,Jun:15} have similar architecture.} We assume that the storage system is arranged as an array. We refer to the $j$-th server as the $j$-th column, and the set of $i$-th storage nodes across all the servers as the $i$-th row of the storage array. Given two positive integers $\delta$ and $d$ such that $\delta < d$, our goal is to encode the data in such a way that 
\begin{enumerate}
\item any crisscross failure affecting at most $\delta-1$ rows and/or columns of nodes in a rack should be  {\it `locally'} recoverable by accessing only the nodes on the corresponding rack, and 
\item any crisscross failure that affects at most $d-1$ rows and/or columns of nodes in the system should be recoverable (potentially by accessing all the remaining data). 
\end{enumerate}
Note that the failure patterns of the first kind can occur in several cases. For example, all the nodes on a server would fail if, say, the network switch connecting the server to the system fails. The entire row of nodes might be temporarily unavailable in certain scenarios, for instance, if these nodes are simultaneously scheduled for an upgrade. A few locally recoverable crisscross patterns are shown in Fig.~\ref{fig:9by9-array-erasures} (considering $\delta = 3$). Note that locally recoverable erasures in different racks can be simultaneously repaired.

\end{example}

Next, we extend the notion of locality from the rank metric to the subspace distance metric. Let $\GF{q}^M$ denote the vector space of $M$-tuples over $\GF{q}$.  A subspace code is a non-empty set of subspaces of $\GF{q}^{M}$. A subspace code in which each codeword has the same dimension is called a constant-dimension code {or a Grassmannian code} (see, \eg,~\cite{KoetterK:08,Khaleghi:09}). A useful distance measure between two spaces $U$ and $V$, called {\it subspace metric}, is defined in~\cite{KoetterK:08} as $\ds{U,V} =  \dims{U} + \dims{V} - 2\:\dims{U\cap V}$. 
{To define the notion of subspace-locality, we need to to choose an ordered basis for every codeword subspace.}
{For a Grassmannian code, we say that the $i$-th basis vector has $(r,\delta)$ {\it subspace-locality}, if there exists a set $\gam{i}$  of basis vectors of size at most $r+\delta-1$ such that $\gam{i}$ contains $i$ and the code obtained by removing the basis vectors outside $\gam{i}$ for each codeword has subspace distance at least $\delta$.} We say that a {Grassmannian code} has $(r,\delta)$ {subspace-locality} if every basis vector has $(r,\delta)$ {subspace-locality}.

{Grassmannian codes} play an important role in correcting errors and erasures (rank-deficiencies) in non-coherent linear network coding~\cite{KoetterK:08,SilvaK:09}. We present an application of the proposed novel Grassmannian codes with locality for downloading partial data and repairing failed nodes in a distributed storage system, in which the nodes are connected over a network that can introduce errors and erasures. The locality is useful when a user wants to download partial data by connecting to only a small subset of nodes, or while repairing a failed storage node over the network (see Sec.~\ref{sec:networked-storage}).

{\bf Our Contributions:}
First, we introduce the notion of locality in rank metric. Then, we establish a tight upper bound on the minimum rank distance of codes with $(r,\delta)$ rank-locality. We construct a family of {\it optimal} codes which achieve this upper bound. Our approach is inspired by the seminal work by Tamo and Barg~\cite{TamoB:14}, which generalizes Reed-Solomon code construction to obtain codes with locality. We generalize the Gabidulin code construction~\cite{Gabidulin:85} to design codes with rank-locality. In particular, we obtain codes as evaluations of specially constructed {\it linearized polynomials} over an extension field, and our codes reduce to Gabidulin codes if the locality parameter $r$ equals the code dimension. We also characterize various erasure and error patterns that the proposed codes with rank-locality can efficiently correct. 

Second, we extend the notion of locality to the subspace metric. Then, we consider a method to construct {Grassmannian codes} by {\it lifting} rank-metric codes (proposed by Silva \etal~\cite{SilvaKK:08}), and show that a  {Grassmannian code} obtained by lifting an array code with rank-locality possesses subspace-locality. This enables us to construct a novel family of  {Grassmannian codes} with subspace-locality by lifting the proposed rank-metric codes with rank-locality. Finally, we highlight an application of codes with subspace-locality in networked distributed storage systems.

\section{Preliminaries}
\label{sec:basics}

\subsection{Notation}
\label{sec:notation}
We use the following notation.
For an integer $l$, $[l] = \{1,2,\ldots,l\}$.
For a vector $\vect{x}$, $\wt{\vect{x}}$ denotes its Hamming weight, \ie, $\wt{\vect{x}} = |\{i : \vect{x}(i) \neq 0\}|$.
The transpose, rank and column space of a matrix $H$ is denoted by $H^T$, $\rnk{H}$, and $\left\langle H \right\rangle$, respectively. The linear span of a set of vectors $\vect{x}_1, \ldots, \vect{x}_k$ is denoted by $\left\langle \vect{x}_1, \ldots, \vect{x}_k\right\rangle$. {We define the reduced column echelon form (RCEF) of a matrix $H$, denoted by $\rcef{H}$, as the transpose of the reduced row echelon form of $H^T$. In other words, one first performs row operations on $H^T$ to transform it to the reduced row echelon form, and then takes its transpose to obtain $\rcef{H}$.}

Let $\code$ denote a linear $(n,k)$ code over $\GF{q}$ with block-length $n$, dimension $k$, and minimum distance $\dm{\code}$. For instance, under Hamming metric, we have $\dm{\code} = \min_{\cwset{i},\cwset{j}\in\code, \: \cwset{i}\neq\cwset{j}}\wt{\cwset{i}-\cwset{j}}$. Given a length-$n$ block code $\code$ and a set $\mathcal{S}\subset[n]$, let $\codeproj{\mathcal{S}}$ denote the restriction of $\code$ on the coordinates in $\mathcal{S}$. Equivalently, $\codeproj{\mathcal{S}}$ is the code obtained by puncturing $\code$ on $[n]\setminus\mathcal{S}$.

Recall that, for Hamming metric, the well known Singleton bound gives an upper bound on the minimum distance of an $(n,k)$ code $\code$ as $\dm{\code} \leq n - k + 1$. Codes which meet the Singleton bound are called maximum distance separable (MDS) codes (see, \eg,~\cite{MacWilliamsS:77}).

\subsection{Codes with Locality}
\label{sec:LRCs}

Locality of a code captures the number of symbols participating in recovering a lost symbol. In particular, an $(n,k)$ code is said to have locality $r$ if every symbol is recoverable from a set of at most $r$ other symbols. For linear codes with locality, a {\it local} parity check code of length at most $r+1$ is associated with every symbol. 
The notion of locality can be generalized  to accommodate {\it local codes} of larger distance as follows (see~\cite{Prakash:12}).

\begin{definition}[Locality]
\label{def:locality}
An $(n, k)$ code $\code$ is said to have $(r,\delta)$ locality, if for every coordinate $i\in[n]$, 
there exists a set of indices $\gam{i}$ such that
\begin{enumerate}
\item $i\in\gam{i}$,
\item $|\gam{i}| \leq r + \delta - 1$, and
\item $\dm{\codeproj{\gam{i}}} \geq \delta$.
\end{enumerate}
The code $\codeproj{\gam{i}}$ is said to be the local code associated with the $i$-th coordinate of $\code$.
\end{definition}
Properties 2 and 3 imply that for any codeword in $\code$, the values in $\gam{i}$ are uniquely determined by any $r$ of those values. Under Hamming metric, the $(r,\delta)$ locality allows one to {\it locally} repair any $\delta-1$ erasures in $\codeproj{\gam{i}}$, $\forall i\in[n]$, by accessing at most $r$ other symbols. When $\delta = 2$, the above definition reduces to the classical definition of locality proposed by Gopalan \etal~\cite{Gopalan:12}, wherein any one erasure can be repaired by accessing at most $r$ other symbols. 

The Singleton bound can be generalized to accommodate locality constraints. In particular,  the minimum Hamming distance of an $(n,k)$ code $\code$ with $(r,\delta)$ locality is upper bounded as follows (see~\cite[Theorem 21]{Rawat:14}, also~\cite[Theorem 2]{Prakash:12} for linear codes):
\begin{equation}
\label{eq:dist-bound-LRC}
\dm{\code} \leq n - k + 1 - \left(\ceillr{\frac{k}{r}} - 1\right)(\delta - 1).
\end{equation}

\section{Codes with Rank-Locality}
\label{sec:local-MRD}

 \subsection{Rank-Metric Codes}
\label{sec:rank-metric-codes}
Let $\GFmn{q}$ be the set of all $m\times n$ matrices over $\GF{q}$. The {\it rank distance} is a distance measure between elements $A$ and $B$ of $\GFmn{q}$, defined as $\dr{A,B} = \rnk{A-B}$. It can be shown that the rank distance is indeed a metric~\cite{Gabidulin:85}. A rank-metric code is a non-empty subset of $\GFmn{q}$ equipped with the rank distance metric (see~\cite{Delsarte:78, Gabidulin:85, Roth:91}). Rank-metric codes can be considered as array codes or matrix codes. 

The minimum rank distance of a code $\code$ is given as $$\dr{\code} = \min_{C_i,\: C_j\in\code,\:\: C_i\ne C_j} \dr{C_i,C_j}.$$ 
We refer to a linear code $\code\subset\GFmn{q}$ with cardinality $|\code| = (q^m)^k$ and minimum rank distance $d$ as an $(m\times n, k, d)$ code.

The Singleton bound for the rank metric (see~\cite{Gabidulin:85}) states that every rank-metric code with minimum rank distance $d$ must satisfy 
$$|\code| \leq q^{\max\{n,m\}(\min\{n,m\} - d + 1)}.$$ 
Codes that achieve this bound are called maximum rank distance (MRD) codes.

A minimum distance decoder for a rank-metric code $\code\subseteq\GFmn{q}$ takes an array $Y\in\GFmn{q}$ and returns a codeword $X\in\code$ that is closest to $Y$ in rank distance. In other words, 
\begin{equation}
\label{eq:min-rank-distance-decoder}
X = \arg\min_{X'\in\code} \rnk{Y - X'}.
\end{equation}

Typically, rank-metric codes are considered by leveraging the correspondence between $\GF{q}^{m\times 1}$ and the extension field $\GFm{q}$ of $\GF{q}$. In particular, by fixing a basis for $\GFm{q}$ as an $m$-dimensional vector space over $\GF{q}$, any element of $\GFm{q}$ can be represented as a length-$m$ vector over $\GF{q}$. Similarly, any length-$n$ vector over $\GFm{q}$ can be represented as an $m\times n$ matrix over $\GF{q}$. The rank of a vector $\vect{a}\in\GFm{q}^n$ is the rank of the corresponding $m\times n$ matrix $A$ over $\GF{q}$. This rank does not depend on the choice of basis for $\GFm{q}$ over $\GF{q}$. This correspondence allows us to view a rank-metric code in $\GFmn{q}$ as a block code of length $n$ over $\GFm{q}$. 
{Further, when viewed as a block code over $\GFm{q}$, an $(m\times n,k,d)$ MRD code (over $\GF{q}$) is  an $[n,k,d]$ MDS code (over $\GFm{q}$), and hence can correct any $n-k$ {\it column erasures}.}

Gabidulin~\cite{Gabidulin:85} presented a construction of a class of MRD codes for $m\geq n$. The construction is based on the evaluation of a special type of polynomials called {\it linearized polynomials}. {We present a brief review of linearized polynomials and Gabidulin construction in Appendix~\ref{app:lin-poly}.}

\subsection{Locality in the Rank Metric}
\label{sec:LRCs}

Recall from Definition~\ref{def:locality} that, for a code $\code$ with $(r,\delta)$ locality, the {\it local code} $\codeproj{\gam{i}}$ associated with the $i$-th symbol, $i\in[n]$, has minimum distance at least $\delta$. 
We are interested in rank-metric codes such that the local code associated with every column should be a rank-metric code with minimum rank distance guarantee. This motivates us to generalize the concept of locality to that of {\it rank-locality} as follows.

\begin{definition}[Rank-Locality]
\label{def:local-rank-metric-code}
An $(m\times n, k)$ rank-metric code $\code$ is said to have $(r,\delta)$ rank-locality, if for every column $i\in[n]$, 
there exists a set of columns $\gam{i}\subset [n]$ such that
\begin{enumerate}
\item $i\in\gam{i}$,
\item $|\gam{i}| \leq r + \delta - 1$, and
\item $\dr{\codeproj{\gam{i}}} \geq \delta$,
\end{enumerate}
where $\codeproj{\gam{i}}$ is the restriction of $\code$ on the columns indexed by $\gam{i}$. The code $\codeproj{\gam{i}}$ is said to be the local code associated with the $i$-th column. An $(m\times n, k)$ rank-metric code with minimum distance $d$ and $(r,\delta)$ locality is denoted as an $(m\times n, k, d, r, \delta)$ rank-metric code.
\end{definition}

As we will see in Section~\ref{sec:erasure-correction-capability}, the $(r,\delta)$-rank-locality allows us to repair any crisscross erasure pattern of {\it weight} $\delta - 1$ in $\codeproj{\gam{i}}$, $\forall i\in[n]$, {\it locally} by accessing the symbols of $\codeproj{\gam{i}}$. 

\subsection{Upper Bound on Rank Distance}
It is easy to find the Singleton-like upper bound on the minimum rank distance for codes with rank-locality using the results in the Hamming metric. 

\begin{theorem}
\label{thm:dist-bound-rank-locality}
For a rank-metric code $\code\subseteq\GFmn{q}$ of cardinality $q^{mk}$ with $(r,\delta)$ rank-locality, we have 
\begin{equation}
\label{eq:dist-bound-rank-locality} 
\dr{\code} \leq n - k + 1 - \left(\ceillr{\frac{k}{r}} - 1 \right)(\delta - 1).
\end{equation}
\end{theorem} 
\begin{IEEEproof}
Note that by fixing a basis for $\GFm{q}$ as a vector space over $\GF{q}$, we can obtain a bijection $\phi : \GFm{q}\rightarrow \GF{q}^{m\times 1}$. This can be extended to a bijection $\phi : \GFm{q}^n \rightarrow \GF{q}^{m\times n}$. Then, for any vector $\cw\in\GFm{q}^n$, there is a corresponding matrix $C\in\GF{q}^{m\times n}$ such that $C=\phi(\cw)$. For any such vector-matrix pair, we have 
\begin{equation}
\label{eq:rank-vs-dist}
\rnk{C}\leq\wt{\cw}.
\end{equation}

An $(m\times n, k, d)$ rank-metric code $\code$ over $\GF{q}$ can be considered as a block code of length $n$ over $\GFm{q}$, denoted as $\code'$. From~\eqref{eq:rank-vs-dist}, it follows that $\dr{\code}\leq\dm{\code'}$. Moreover, it follows that, if $\code$ has $(r,\delta)$ rank-locality, then the corresponding code $\code'$ possesses $(r,\delta)$ locality in  the Hamming metric. Therefore, an upper bound on the minimum Hamming distance of an $(n,k,d')$-LRC $\code'$ with $(r,\delta)$ locality is also an upper bound on the rank distance of an $(m\times n,k,d)$ rank-metric code with $(r,\delta)$ rank-locality. Hence,~\eqref{eq:dist-bound-rank-locality} follows from~\eqref{eq:dist-bound-LRC}.
\end{IEEEproof}

\section{A Class of Optimal Codes with Rank-Locality}
\label{sec:code-construction}

\subsection{Code Construction}
\label{sec:code-construction-subsection}
We build upon the construction methodology of Tamo and Barg~\cite{TamoB:14} to construct codes with rank-locality that are optimal with respect to the rank distance bound in~\eqref{eq:dist-bound-rank-locality}.\footnote{We present a detailed comparison of our construction with that of~\cite{TamoB:14} in Sec.~\ref{rem:comparison-Tamo-Barg}.} In particular, the codes are constructed as the evaluations of specially designed linearized polynomials\footnote{We refer the reader to Appendix~\ref{app:lin-poly} for a brief review of linearized polynomials and Gabidulin code construction.} on a specifically chosen set of points of $\GFext{q}{m}$. The detailed construction is as follows. For notational convenience, we write $x^{q^i} = \xq{x}{i}$.

\begin{construction}[$(m\times n, k, r, \delta)$ rank-metric code]
\label{con:local-MRD-1}
Let $m, n, k, r,$ and  $\delta$ be positive integers such that $r\mid k$, $(r+\delta-1)\mid n$, and $n\mid m$. Define $\mu := {n}/{(r+\delta-1)}$. Fix $q\geq 2$ to be a power of a prime. 
Let $\basis{A} = \left\{\al{1},\ldots,\al{r+\delta-1}\right\}$ be a basis of $\GFext{q}{r+\delta-1}$ as a vector space over $\GF{q}$, and $\basis{B} = \left\{\be{1},\ldots,\be{\mu}\right\}$ be a basis of $\GFext{q}{n}$ as a vector space over $\GFext{q}{r+\delta-1}$. Define the set of $n$ evaluation points 
$\setP = \setP_{1} \cup \cdots \cup \setP_{\mu}$, where $\setP_{j} = \left\{\al{i}\be{j}, 1\leq i\leq r+\delta-1\right\}$ for $1\leq j\leq \mu$. To encode the message $\m\in\GFext{q}{m}^k$, denoted as $\m = \left\{m_{ij} : i = 0,\ldots, r-1; j = 0,\ldots,\frac{k}{r}-1\right\}$, define the encoding polynomial 
\begin{equation}
\label{eq:gen-poly}
G_{\m}(x) = \sum_{i=0}^{r-1}\sum_{j=0}^{\frac{k}{r}-1}\mij{ij}\xq{x}{(r+\delta-1)j+i}.
\end{equation}
The codeword for $\m$ is obtained as the vector of the evaluations of $G_{\m}(x)$ at all the points of $\setP$. In other words, the linear code $\code_{Loc}$ is constructed as the following evaluation map:
\begin{IEEEeqnarray}{rCl}
\label{eq:eval-map}
Enc & : & \GFext{q}{m}^k \rightarrow \GFext{q}{m}^n\nonumber\\
& & \m \mapsto \left\{G_{\m}(\gamma), \gamma\in\setP\right\}.
\end{IEEEeqnarray}
Therefore, we have 
\begin{equation}
\label{eq:eval-map-2}
\code_{Loc} = \left\{\left(G_{\m}(\gamma),\gamma\in\setP\right) \mid \m\in\GFext{q}{m}^{k}\right\}.
\end{equation}
The $(m\times n, k)$ rank-metric code is obtained by considering the matrix representation of every codeword obtained as above by fixing a basis of $\GFext{q}{m}$ over $\GF{q}$. We denote the following $\mu$ codes as the local codes.
\begin{equation}
\label{eq:eval-map-local-codes}
\code_{j} = \left\{\left(G_{\m}(\gamma),\gamma\in\setP_j\right) \mid \m\in\GFext{q}{m}^{k}\right\}, \quad 1\leq j\leq \mu.
\end{equation}
\end{construction}

\begin{remark}[Field Size]
\label{rem:field-size}
It is worth mentioning that, as in the construction of Gabidulin codes of length $n$ over $\GFext{q}{m}$~\cite{Gabidulin:85}, it is required that $m\geq n$. Note that, it is sufficient to choose $m = n$ and $q = 2$ in our construction. {In other words, when considered as a block code of length-$n$, the field size of $2^n$ is sufficient for the proposed code construction.} 
\end{remark}

{In the following, we show that Construction~\ref{con:local-MRD-1} gives codes with rank-locality, which are optimal with respect to the rank distance bound in Theorem~\ref{thm:dist-bound-rank-locality}. In the proof, we use some properties of linearized polynomials which are listed in Appendix~\ref{app:lin-poly}. We begin with the two key lemmas that will be used in the proof. The following lemma will be used to prove the rank distance optimality.}

\begin{lemma}
\label{lem:linear-indep}
The $n$ evaluation points given in Construction~\ref{con:local-MRD-1}, $\setP = \left\{\al{i}\be{j}, 1\leq i\leq r+\delta-1, 1\leq j \leq \mu\right\}$, are linearly independent over $\GF{q}$.
\end{lemma}
\begin{IEEEproof}
Suppose, for contradiction, that the evaluation points are linearly dependent over $\GF{q}$. Then, we have $\sum_{j=1}^{\mu}\sum_{i=1}^{r+\delta-1}\wij\al{i}\be{j} = 0$ with coefficients $\wij\in\GF{q}$ such that not all $\omega_{ij}$'s are zero. We can write the linear dependence condition as $\sum_{j=1}^{\mu}\left(\sum_{i=1}^{r+\delta-1}\wij\al{i}\right)\be{j} = 0$. Now, from the linear independence of the $\be{j}$'s over $\GFext{q}{r+\delta-1}$, we have $\sum_{i=1}^{r+\delta-1}\wij\al{i} = 0$ for each $1\leq j\leq \mu$. However, as the $\al{i}$'s are linearly independent over $\GF{q}$, we have every $\wij = 0$. This is a contradiction.  
\end{IEEEproof}

Next, we present a lemma that will be used to prove the rank-locality for the proposed construction. 
Towards this, define $H(x) = x^{q^{r+\delta-1}-1} = x^{[r+\delta-1]-1}$. 
We note that~\eqref{eq:gen-poly} can be written in the following form using $H(x)$:
\begin{equation}
\label{eq:gen-poly-2}
G_{\m}(x) = \sum_{i=0}^{r-1}G_i(x)\xq{x}{i},
\end{equation}
where 
\begin{equation}
\label{eq:G-i-x}
G_i(x) = \mij{i0} + \sum_{j=1}^{\frac{k}{r}-1}\mij{ij}[H(x)]^{\sum_{l=0}^{j-1}q^{(r+\delta-1)l + i}}.
\end{equation}
To see this, observe that
\begin{IEEEeqnarray}{rCl}
[H(x)]^{\sum_{l=0}^{j-1}q^{(r+\delta-1)l + i}} &=& \left[x^{q^{r+\delta-1}-1}\right]^{\sum_{l=0}^{j-1}q^{(r+\delta-1)l + i}}\nonumber\\
&=& x^{\sum_{l=0}^{j-1}q^{(r+\delta-1)(l+1)+i} - \sum_{l=0}^{j-1}q^{(r+\delta-1)l+i}}\nonumber\\
\label{eq:H-x-alternative}
&=& x^{q^{(r+\delta-1)j+i}-q^i}.
\end{IEEEeqnarray}
Now, using~\eqref{eq:H-x-alternative} in~\eqref{eq:G-i-x}, we get
\begin{equation}
\label{eq:G-i-x-2}
G_i(x) = \mij{i0} + \sum_{j=1}^{\frac{k}{r}-1}\mij{ij} x^{[(r+\delta-1)j+i] - [i]}.
\end{equation}
Then, substituting~\eqref{eq:G-i-x-2} into~\eqref{eq:gen-poly-2} gives us~\eqref{eq:gen-poly}.

Next, we prove that $H(x)$ is constant on all points of $\setP_j$ for each $1\leq j\leq \mu$.
\begin{lemma}
\label{lem:hx-good-poly}
Consider the partition of the set of evaluation points given in Construction~\ref{con:local-MRD-1} as $\setP = \setP_1 \cup \cdots \cup \setP_{\mu}$, where $\setP_j =  \left\{\al{i}\be{j}, 1\leq i\leq r+\delta-1\right\}$. Then, $H(x)$ is constant on all evaluation points of any set $\setP_j$ for $1\leq j \leq \mu$.
\end{lemma}
\begin{IEEEproof}
Note that $H(\be{j}\al{i}) = \left(\be{j}\al{i}\right)^{[r+\delta-1]-1} = \be{j}^{[r+\delta-1]-1}\al{i}^{[r+\delta-1]-1} = \be{j}^{[r+\delta-1]-1}$, where the last equality follows from $\al{i}\in\GFext{q}{r+\delta-1}\setminus\{0\}$. Thus, $H(\omega) = \be{j}^{[r+\delta-1]-1}$, for all $\omega\in\setP_j$, $1\leq j \leq \mu$.
\end{IEEEproof}

{Now, we use Lemmas~\ref{lem:linear-indep} and~\ref{lem:hx-good-poly} to prove the rank-locality and rank distance optimality of the proposed construction.}

\begin{theorem}
\label{thm:optimality-of-construction-1}
Construction~\ref{con:local-MRD-1} gives a linear $(m\times n,k,d)$ rank-metric code $\code_{Loc}$ with $(r,\delta)$ rank-locality such that the minimum rank distance $d$ is equal to the upper bound given in~\eqref{eq:dist-bound-rank-locality}.
\end{theorem}
\begin{IEEEproof}
We begin with showing the rank distance optimality of $\code_{Loc}$.
{Lemma~\ref{lem:linear-indep} asserts} that $\code_{Loc}$ is obtained as the evaluations of $G_{\m}(x)$ on $n$ points of $\GFext{q}{m}$ that are linearly independent over $\GF{q}$.  Combining this with the structure of $G_{\m}(x)$ (see~\eqref{eq:gen-poly}), $\code_{Loc}$ can be considered as a subcode of an $\left(n, k+\left(\frac{k}{r}-1\right)(\delta-1) \right)$ Gabidulin code ({\it cf.}~\eqref{eq:Gabidulin-lin-poly} in Appendix~\ref{app:lin-poly}). Hence, $\dr{\code_{Loc}} \geq n - k + 1 -\left(\frac{k}{r}-1\right)(\delta-1)$. This shows that $\dr{\code_{Loc}}$ attains the upper bound~\eqref{eq:dist-bound-rank-locality} in Theorem~\ref{thm:dist-bound-rank-locality}, and thus, the proposed construction is optimal with respect to rank distance.\footnote{In Appendix~\ref{app:dist-optimality-proof}, we present an alternative proof from first principles using the properties of linearized polynomials.} 

Second, we show that $\code_{Loc}$ has $(r,\delta)$ rank-locality. 
Towards this, we want to show that $\dr{\code_j}\geq\delta$ for every local code $\code_j$, $1\leq j\leq \mu$.  Let $\gamma \in P_j$ and define the {\it repair polynomial} as
\begin{equation} 
\label{eq:rep-poly}
R_j(x) = \sum_{i=0}^{r-1}G_i(\gamma)\xq{x}{i},
\end{equation}
where $G_i(\cdot)$ is defined in~\eqref{eq:G-i-x}.
We show that $\code_j$ can be considered as obtained by evaluating $R_j(x)$ on the points of $\setP_j$. 

From~\eqref{eq:G-i-x}, observe that $G_i(x)$ is a linear combination of powers of $H(x)$. From Lemma~\ref{lem:hx-good-poly}, $H(x)$ is constant on $\setP_j$. Therefore, $G_i
(x)$ is also constant on $\setP_j$. In other words, we have
\begin{equation}
\label{eq:G-i-const}
G_i(\gamma) = G_i(\lambda), \quad \forall\: \gamma, \lambda\in\setP_j,
\end{equation}
for every $0\leq i\leq r-1$.

Moreover, when evaluating $R_j(x)$ in $\lambda \in P_j$, we get
\begin{equation}
\label{eq:R-x-eq-G-x}
R_j(\lambda) = \sum_{i=0}^{r-1}G_i(\gamma)\xq{\lambda}{i} = \sum_{i=0}^{r-1}G_i(\lambda)\xq{\lambda}{i} = G_{\m}(\lambda).
\end{equation}
Hence, the evaluations of the encoding polynomial $G_{\m}(x)$ and the repair polynomial $R_j(x)$ on points in $\setP_j$ are identical. Therefore, we can consider that $\code_j$ is obtained by evaluating $R_j(x)$ on points of $\setP_j$. Now, since points of $\setP_j$ are linearly independent over $\GF{q}$, and $R_j(x)$ is a linearized polynomial of $q$-degree $r-1$, $\code_j$ can be considered as a $(r+\delta-1,r)$ Gabidulin code ({\it cf.}~\eqref{eq:Gabidulin-lin-poly} in Appendix~\ref{app:lin-poly}). Thus, $\code_j$ is an MRD code, and we have $\dr{\code_j} = \delta$, which proves the rank-locality of the proposed construction.\footnote{We note that the result $\dr{\code}\geq \delta$ also follows from Lemma~\ref{lem:lin-poly-encoding} in Appendix~\ref{app:dist-optimality-proof}, which is proved from first principles using the properties of linearized polynomials.} 
\end{IEEEproof}

{We note that, in Construction~\ref{con:local-MRD-1}, we assume that $(r+\delta-1)\mid n$. Generalizing the construction when $(r+\delta-1)\nmid n$ does not seem to be straightforward, and it is left as a future work.}

Next, we present an example of an $(9\times 9, 4)$ rank-metric code with $(2, 2)$ rank-locality. We note that the code presented in this example satisfies the correctability constraints specified in the motivating example (Example~\ref{ex:9by9-array}) in the Introduction section.
\begin{example}
\label{ex:k-4-r-2-code}
Let $n = 9, k = 4, r = 2, \delta = 2$. Set $q = 2$ and $m=n$. 
Let $\omega$ be the primitive element of $\GFext{2}{9}$ with respect to the primitive polynomial $p(x) = x^9 + x^4 + 1$. 
Note that $\omega^{73}$ generates $\GFext{2}{3}$, as $\left(\omega^{73}\right)^7 = 1$. Consider $\basis{A} = \{1, \omega^{73}, \omega^{146}\}$ as a basis for $\GFext{2}{3}$ over $\GF{2}$. We view $\GFext{2}{9}$ as an extension field over $\GFext{2}{3}$ considering the irreducible polynomial $p(x) = x^3 + x + \omega^{73}$. It is easy to verify that $\omega^{309}$ is a root of $p(x)$, and thus, $\basis{B} = \{1, \omega^{309}, \omega^{107}\}$ forms a basis of $\GFext{2}{9}$ over $\GFext{2}{3}$. Then, the evaluation points $\setP$ and their partition $\setPpartition$ is as follows. 
\begin{IEEEeqnarray}{rCl}
\setPpartition  & = &  
\left\{
\setP_1 = \{1, \omega^{73}, \omega^{146}\},  
\setP_2 = \{\omega^{309}, \omega^{382}, \omega^{455}\},
\right. 
\nonumber\\ 
&  & 
\:\: \left.
\setP_3 =\{\omega^{107}, \omega^{180}, \omega^{253}\}
\right\}.\nonumber
\end{IEEEeqnarray}

Let $\m = (\mij{00},\: \mij{01},\: \mij{10},\: \mij{11}) \in \GFext{2}{9}^4$ be the information vector. Define the encoding polynomial (as in~\eqref{eq:gen-poly}) as follows.
$$G_{\m}(x) = \mij{00}\xq{x}{0} + \mij{01}\xq{x}{3} + \mij{10}\xq{x}{1} + \mij{11}\xq{x}{4}.$$ 
The codeword $\cw$ for the information vector $\m$ is obtained as the evaluation of the polynomial $G_{\m}(x)$ at all the points of $\setP$. The code $\code$ is the set of  codewords corresponding to all $\m\in\GFext{2}{9}^4$.

From Lemma~\ref{lem:linear-indep}, the evaluation points are linearly independent over $\GF{2}$, and thus, $\code$ can be considered as a subcode of a $(9,5)$ Gabidulin code (cf.~\eqref{eq:Gabidulin-lin-poly}). Thus, $\dr{\code} = 5$, which is optimal with respect to~\eqref{eq:dist-bound-rank-locality}.

Now, consider the local codes $\code_j$, $1\leq j\leq 3$. It is easy to verify that $\code_j$ can be  obtained by evaluating the {\it repair polynomial} $R_j(x)$ on $\setP_j$ given as follows (see~\eqref{eq:rep-poly}).
\begin{IEEEeqnarray}{rCl}
R_1(x) & = & (\mij{00} + \mij{01})\xq{x}{0} + (\mij{10} + \mij{11})\xq{x}{1},\nonumber\\
R_2(x) & = & (\mij{00} + \omega^{119}\mij{01})\xq{x}{0} + (\mij{10} + \omega^{238}\mij{11})\xq{x}{1},\nonumber\\
R_3(x) & = & (\mij{00} + \omega^{238}\mij{01})\xq{x}{0} + (\mij{10} + \omega^{476}\mij{11})\xq{x}{1}\nonumber.
\end{IEEEeqnarray}
For instance, let the message vector be $\vect{m} = (\omega, \omega^2, \omega^4, \omega^8)$. Then, the codeword is $$\cw = (\omega^{440},\omega^{307},\omega^{81},\omega^{465},\omega^{11},\omega^{174},\omega^{236},\omega^{132},\omega^{399}).$$ One can easily check that evaluating $R_1(x)$ on $\setP_1$ gives $\cw_1 = (\omega^{440},\omega^{307},\omega^{81})$, evaluating $R_2(x)$ on $\setP_2$ gives $\cw_2 = (\omega^{465},\omega^{11},\omega^{174})$, and evaluating $R_3(x)$ on $\setP_3$ gives $\cw_3 = (\omega^{236},\omega^{132},\omega^{399})$.

This implies that the local code $\code_j$, $1\leq j\leq 3$, can be considered as obtained by evaluating a linearized polynomial of the form $R_j(x) = \mij{0}'\xq{x}{0} + \mij{1}'\xq{x}{1}$ on three points that are linearly independent over $\GF{2}$. Hence, $\code_j$ is a Gabidulin code of length 3 and dimension 2, which gives 
$\dr{\code_j} = 2$. This shows that $\code$ has $(2,2)$ rank-locality. 
\end{example}
 
\subsection{Comparison with Tamo and Barg~\cite{TamoB:14}}
\label{rem:comparison-Tamo-Barg}
The key idea in~\cite{TamoB:14} is to construct codes with locality as evaluations of a specially designed polynomial over a specifically chosen set of elements of the underlying finite field. To point out the similarities and differences, we briefly review Construction 8 from~\cite{TamoB:14}. We assume that $r\mid k$, and $(r+\delta-1)\mid n$.

{\bf Construction 8 from~\cite{TamoB:14}:}  Let $\setPpartition =\{\setP_1, \ldots,\setP_{\mu}\}$, $\mu=n/(r+\delta-1)$, be a partition of the set $\setP \subset \GF{q}$, $|\setP| = n$, such that $| \setP_i | = r + \delta - 1$, $1\leq i \leq \mu$. Let $h \in \GF{q}[x]$ be a polynomial of degree $r + \delta - 1$, called the {\it good polynomial}, that is constant on each of the sets $\setP_i$. For an information vector $\m\in\GF{q}^k$, define the encoding polynomial
$$g_{\m}(x) = \sum_{i=0}^{r-1}\left(\sum_{j=0}^{\frac{k}{r}-1}\mij{ij}h(x)^j\right)x^i.$$
The code $\code$ is defined as the set of $n$-dimensional vectors
$$\code = \left\{\left(g_{\m}(\gamma),\gamma\in\setP\right) \mid \m\in\GF{q}^{k}\right\}.$$

The authors show that $h(x) = x^{r+\delta-1}$ can be used as a {\it good polynomial}, when the evaluation points are cosets of a multiplicative subgroup of $\GF{q}^{*}$ of order $r+\delta-1$.
In this case, we can write $g_{\m}(x)$ as
\begin{equation}
\label{eq:Tamo-Barg}
g_{\m}(x) = \sum_{i = 0}^{r-1}\sum_{j=0}^{\frac{k}{r}-1}\mij{ij}x^{(r+\delta-1)j + i}. 
\end{equation}
Therefore, $\code$ can be considered as a subcode of an $\left(n, k+\left(\frac{k}{r}-1\right)(\delta-1) \right)$ Reed-Solomon code. In addition, local codes $\code_j =\left\{\left(g_{\m}(\gamma),\gamma\in\setP_j\right) \mid \m\in\GF{q}^{k}\right\}$, $1\leq j\leq \mu$, can be considered as $(r+\delta-1,r)$ Reed-Solomon codes.

In our case, the code $\code_{Loc}$ obtained from Construction~\ref{con:local-MRD-1} can be considered as a subcode of a $\left(n, k+\left(\frac{k}{r}-1\right)(\delta-1) \right)$ Gabidulin code. Further, the local codes $\code_j$, $1\leq j\leq \mu$, can be considered as $(r+\delta-1,r)$ Gabidulin codes. In fact, as one can see from the proof of Theorem~\ref{thm:optimality-of-construction-1}, we implicitly use $H(x) = x^{[r+\delta-1]-1}$ as the good polynomial, which evaluates as a constant on all points of $\setP_j$ for $1\leq j\leq \mu$ given in Construction~\ref{con:local-MRD-1}. It is worth mentioning that \eqref{eq:Tamo-Barg} and \eqref{eq:gen-poly} turn out to be $q$-associates of each other; see Definition~\ref{def:q-associates} in Appendix~\ref{app:lin-poly}.

\subsection{Comparison with Silberstein \etal~~\cite{Silberstein:13}} 
\label{rem:comparison-Silberstein}
In~\cite{Silberstein:13} (see also~\cite{Rawat:14}), the authors  have presented a construction of LRC codes based on rank-metric codes. The idea is to first precode the information vector with an $(r\mu,k)$ Gabidulin code over $\GFext{q}{m}$. The symbols of the codeword are then partitioned into $\mu$ sets $C_1, \ldots, C_{\mu}$ of size $r$ each. For each set $C_j$, an $(r+\delta-1,r)$ Reed-Solomon code over $\GF{q}$ is used to obtain $\delta-1$ local parities, which together with the symbols of $C_j$ form the codeword of a local code $\code_j$. This ensures that each local code has minimum distance $\delta$. 
However, it does not guarantee that the minimum rank distance of a local code is at least $\delta$. 

In fact, for any $\cw\in\code_j$, $1\leq j\leq \mu$, 
we have $\rnk{\cw} \leq r$, as the local parities are obtained via linear combinations over $\GF{q}$. Clearly, when $\delta > r$, the construction cannot achieve rank-locality. Moreover, even if $\delta \leq r$, it is possible to have a codeword $\cw\in\code_j$ such that $\rnk{\code_i} < \delta$ for some local code $\code_j$. Therefore, in general, the construction of~\cite{Silberstein:13}, that uses Gabidulin codes as outer codes, does not guarantee that the codes possess rank-locality. 

On the other hand, our construction can be viewed as a method to design $(n, k)$ linear codes over $\GFm{q}$ with $(r,\delta)$ locality (under the Hamming metric). For the construction in~\cite{Silberstein:13}, the field size of $q^n$ is sufficient for $q \geq r+\delta-1$ when $\delta>2$, while one can choose any $q\geq 2$ when $\delta = 2$. When our construction is used to obtain LRCs, it is sufficient to operate over the field of size $2^n$.

\section{Correction Capability of Codes with Rank-Locality}
\label{sec:erasure-correction-capability}

Suppose the encoded data is stored on an $m\times n$ array $C$ using an $(m\times n,k,d,r,\delta)$ rank-metric code $\code$ over $\GF{q}$. 
Our goal is to characterize the class of (possibly correlated) mixed erasure and error  patterns corresponding to column and row failures of $C$ that $\code$ can correct locally or globally. 

\begin{remark}
\label{rem:local-codes}
{
In this section, we assume that the columns of an $(m\times n,k,r,\delta)$ rank-metric code $\code$ can be partitioned into $\mu := {n}/{(r+\delta-1)}$ disjoint sets $C_1, \ldots, C_\mu$ each of size $r+\delta-1$ such that, for all $i\in C_j$, $\gam{i} = C_j$. In other words, we assume that the local codes associated with the columns have disjoint coordinates. 
Note that the proposed Construction~\ref{con:local-MRD-1} satisfies this assumption. 
}
\end{remark}

We begin with the notion of crisscross weight of an erasure pattern. Let $E = [\eij{i,j}]_{1 \leq i \leq m, 1 \leq j \leq n}$ be an $m\times n$ binary matrix that specifies the location of the erased symbols of $C$, referred to as an erasure matrix. In particular, $\eij{ij} = 1$ if the $(i,j)$-th entry of $C$ is erased, otherwise $\eij{ij} = 0$. For simplicity, we denote the erasure pattern by $E$ itself. We denote by $E(C_j)$ the $r+\delta-1$ 
columns of $E$ corresponding to the  local array $C_j$, and we refer to $E(C_j)$ as the erasure pattern restricted to the local array $C_j$. We first consider the notion of a {\it cover} of $E$, which is used to define the crisscross weight of $E$ (see~\cite{Roth:91}, also~\cite{Blaum:00}).
\begin{definition}[Cover of $E$]
\label{def:cover-of-E}
(\cite{Roth:91}) A cover of an $m\times n$ matrix $E$ is a pair $(X,Y)$ of sets $X\subseteq[m]$, $Y\subseteq[n]$, such that $\eij{ij}\ne 0 \implies \left((i\in X)\:\: \textrm{or} \:\: (j\in Y)\right)$ for all $1\leq i \leq m$, $1\leq j \leq n$. The size of the cover $(X,Y)$ is defined as $|(X,Y)| = |X| + |Y|$. 
\end{definition} 
We define the crisscross weight of an erasure pattern as the crisscross weight of the associated binary matrix $E$ defined as follows.
\begin{definition}[Crisscross weight of $E$]
\label{def:weight-of-E}
(\cite{Roth:91}) The crisscross weight of an erasure pattern $E$ is the minimum size $|(X,Y)|$ over all possible covers $(X,Y)$ of the associated binary matrix $E$. We denote the crisscross weight of $E$ as $\wtc{E}$. 
\end{definition} 
Note that a minimum-size cover of a given matrix $E$ is not always unique. Further, the minimum size of a cover of a binary matrix is equal to the maximum number of 1's that can be chosen in that matrix such that no two are on the same row or column~\cite[Theorem 5.1.4]{Hall:98}.

Let $\Err = [\erij{i,j}]_{1 \leq i \leq m, 1 \leq j \leq n}\in\GFmn{q}$ be a matrix that specifies the location and values of errors occurred in the array, referred to as an error matrix. Specifically,  $\erij{i,j}\in\GF{q}$ denotes the error at the $i$-th row and the $j$-th column. If there is no error, $\erij{i,j} = 0$. We assume that for every $1\leq i\leq m, 1\leq j\leq n$, such that $\eij{i,j} = 1$, we have $\erij{i,j} = 0$. In other words, the value of the error is zero at a location where an erasure occurs. We denote by $\Err(C_j)$ the $r+\delta - 1$ 
columns of $\Err$ corresponding to the  local array $C_j$, and we refer to $\Err(C_j)$ as the error pattern restricted to the local array $C_j$.
 
Now, we characterize erasure and error patterns 
that $\code$ can correct locally or globally.  
Towards this, define a binary variable $\delta_j$ for $1\leq j\leq \mu$ as follows.
\begin{equation}
\label{eq:delta-j}
\delta_j = 
\left\{
\begin{array}{ll}
1 & \textrm{if}\:\: 2\:\rnk{\Err(C_j)} + \wtc{E(C_j)} \leq \delta - 1,\\
0 & \textrm{otherwise}.
\end{array}
\right.
\end{equation}
Recall that, for simplicity, we assume that the local codes associated with columns are disjoint in their support. We note that the proposed construction indeed results in disjoint local codes.

\begin{proposition}
\label{prop:erasure-correction-capability}
Let $\code$ be an $(m\times n, k, d)$ rank-metric code with $(r,\delta)$ rank-locality. Let $\code_j$, $1\leq j\leq \mu$, be the $j$-th local $(r+\delta-1,r,\delta)$ rank-metric code, and let $C_j$ be the corresponding local array. 
Consider erasure and error matrices $E$ and $\Err$. 
{The code $\code_j$ is guaranteed to correct the erasures $E(C_j)$ and errors $\Err(C_j)$ by accessing} the unerased symbols only from $C_j$ provided
\begin{equation}
\label{eq:correction-capability-local}
2\:\rnk{\Err(C_j)} + \wtc{E(C_j)} \leq \delta - 1.
\end{equation} 
Further, the code $\code$ is guaranteed to correct $E$ and $\Err$ provided
\begin{IEEEeqnarray}{l}
\label{eq:correction-capability}
2\:\rnk{\Err} + \wtc{E} \nonumber\\ 
- \sum_{j=1}^{\mu}\delta_j\left(2\:\rnk{\Err(C_j)} + \wtc{E(C_j)}\right)  
\leq d-1,
\end{IEEEeqnarray}
where $\delta_j$ is defined in~\eqref{eq:delta-j}. 
\end{proposition}
\begin{IEEEproof} 
The proof essentially follows from the fact that a rank-metric code $\code$ of rank distance $d$ can correct any erasure pattern $E$ and error pattern $\Err$ such that $2\:\rnk{\Err} + \wtc{E} \leq d-1$. To see this, 
consider a minimum-size cover $(X,Y)$ of $E$. Delete the rows and columns indexed respectively by $X$ and $Y$ in all the codeword matrices of $\code$ as well as from $\Err$ to obtain $\Err'$. The resulting array code composed of matrices of size $m-|X| \times n-|Y|$ has rank distance at least $d - \wtc{E}$. This code can correct any error pattern $\Err'$ such that $\rnk{\Err'} \leq (d - \wtc{E} - 1)/2$ using the minimum distance decoder ({\it cf}.~\eqref{eq:min-rank-distance-decoder}). 
This immediately gives~\eqref{eq:correction-capability-local}. First correcting erasures and errors locally using $\code_j$ for each $1\leq j\leq \mu$, and then globally using $\code$ yields~\eqref{eq:correction-capability}.
\end{IEEEproof} 

\begin{example}
\label{ex:9by9-array-erasures-2}
Suppose the data is to be stored on a $9 \times 9$ bit array $C$ using the $(9\times 9,5,5,2,2)$ rank-metric code discussed in Example~\ref{ex:k-4-r-2-code}. Note that the first three columns of $C$ form the first local array $C_1$, the next three columns form the second local array $C_2$, and the remaining three columns form the third local array $C_3$. The encoding satisfies the correctability constraints mentioned in Example~\ref{ex:9by9-array}. We give an example of the erasure pattern that is correctable in Fig.~\ref{fig:9by9-array-erasures-2}, where locally correctable erasures are denoted as `$?$', while globally correctable erasures are denoted as `$??$'.

\begin{figure}[!t]
\label{fig:9by9-array-erasures-2}
\begin{centering}
\begin{tabular}{|c|c|c| |c|c|c| |c|c|c|}
\hline
$??$ & $??$ & $??$ & 
$??$ & $??$ & $??$ & 
$c_{1,7}$ & $c_{1,8}$ & $c_{1,9}$\\
\hline
$??$ & $??$ & $??$ & 
$??$ & $c_{2,5}$ & $c_{2,6}$ & 
$c_{2,7}$ & $c_{2,8}$ & $c_{2,9}$\\
\hline
$??$ & $??$ & $??$ & 
$??$ & $c_{3,5}$ & $c_{3,6}$ & 
$c_{3,7}$ & $c_{3,8}$ & $c_{3,9}$\\
\hline
$c_{4,1}$ & $c_{4,2}$ & $c_{4,3}$ & 
$??$ & $c_{4,5}$ & $c_{4,6}$ & 
$c_{4,7}$ & $c_{4,8}$ & $c_{4,9}$\\
\hline
$c_{5,1}$ & $c_{5,2}$ & $c_{5,3}$ & 
$??$ & $c_{5,5}$ & $c_{5,6}$ & 
$c_{5,7}$ & $c_{5,8}$ & $c_{5,9}$\\
\hline
$c_{6,1}$ & $c_{6,2}$ & $c_{6,3}$ & 
$??$ & $c_{6,5}$ & $c_{6,6}$ & 
$c_{6,7}$ & $c_{6,8}$ & $c_{6,9}$\\
\hline
$c_{7,1}$ & $c_{7,2}$ & $c_{7,3}$ & 
$??$ & $c_{7,5}$ & $c_{7,6}$ & 
$c_{7,7}$ & $c_{7,8}$ & $c_{7,9}$\\
\hline
$c_{8,1}$ & $c_{8,2}$ & $c_{8,3}$ & 
$??$ & $c_{8,5}$ & $c_{8,6}$ & 
$c_{8,7}$ & $c_{8,8}$ & $c_{8,9}$\\
\hline
$c_{9,1}$ & $c_{9,2}$ & $c_{9,3}$ & 
$??$ & $c_{9,5}$ & $c_{9,6}$ & 
$?$ & $?$ & $?$\\
\hline
\end{tabular}
\caption{An example of a $9\times 9$ bit array. When an erasure pattern affects a single row or column in a local array, it should be corrected locally. Further, any erasure pattern that is confined to at most four rows or columns (or both) should be globally correctable. In the example above, locally correctable erasures are denoted as `$?$', while globally correctable erasures are denoted as `$??$'.}
\end{centering}
\end{figure}

\end{example}
 
\begin{remark} 
\label{rem:discussion}
In Proposition~\ref{prop:erasure-correction-capability}, we only characterize the erasure patterns that are locally or globally correctable. It is interesting to consider efficient decoding algorithms on the lines of~\cite{Gabidulin:08, Silva:09}.  
\end{remark}

\begin{remark} 
\label{rem:discussion-2}
We note that an $(m\times n, k, d, r, \delta)$ code may correct a number of erasure patterns that are not covered by the class mentioned in Proposition~\ref{prop:erasure-correction-capability}. This is analogous to the fact that an LRC can correct a large number of erasures beyond minimum distance. In fact, the class of LRCs that have the maximum erasure correction capability are known as maximally recoverable codes (see~\cite{Gopalan:14}). Along similar lines, it is interesting to extend the notion of maximal recoverability for the rank metric and characterize all the erasure patterns that an $(m\times n, k, d, r, \delta)$ rank-metric code can correct.
\end{remark}

\section{Codes with Subspace-Locality}
\label{sec:local-MSD}

\subsection{Subspace Codes}
\label{sec:subspace-codes}
We briefly review the ideas of subspace codes introduced in~\cite{KoetterK:08}. 
The set of all subspaces of $\GFM{M}$, called the {\it projective space} of order $M$ over $\GF{q}$, is denoted by $\Pq{M}$. The set of all $n$-dimensional subspaces of $\GFM{M}$, called a {\it Grassmannian}, is denoted by $\Gq{M}{n}$, where $0\leq n\leq M$. Note that $\Pq{M} = \cup_{n=0}^{M}\Gq{M}{n}$.

In~\cite{KoetterK:08}, the notion of {\it subspace distance} was introduced. 
Let $U, V \in\Pq{M}$. The {subspace distance} between $U$ and $V$ is defined as
\begin{equation}
\label{eq:subspace-distance}
\ds{U,V} = \dims{U} + \dims{V} - 2\:\dims{U\cap V}.
\end{equation}
It is shown in~\cite{KoetterK:08} that the subspace distance is indeed a metric on $\Pq{M}$. 

A {\it subspace code} is a non-empty subset of $\Pq{M}$ equipped with the subspace distance metric~\cite{KoetterK:08}. The minimum subspace distance of a subspace code $\Omega\subseteq\Pq{M}$ is defined as
\begin{equation}
\label{eq:subspace-min-distance}
\ds{\Omega} = \min_{V_i,V_j\in\Omega,\: V_i\ne V_j} \ds{V_i,V_j}.
\end{equation}

A subspace code $\Omega$ in which each codeword has the same dimension, say $n$, \ie, $\Omega\subseteq\Gq{M}{n}$, is called a {\it constant-dimension code} or a {\it Grassmannian code}. 
It is easy to see, from~\eqref{eq:subspace-distance} and~\eqref{eq:subspace-min-distance}, that the minimum distance of a Grassmannian code is always an even number. {In the rest of the paper, we restrict our attention to Grassmannian codes.} 

{
\begin{remark}
\label{rem:q-analogs}
It is worth noting that several results on subspace codes are $q$-analogs~\cite{Etzion:13} of well-known results on classical codes in the Hamming metric. For instance, Grassmannian codes are $q$-analogs of constant weight codes, and the subspace distance is the $q$-analog of the Hamming distance in the Hamming space. For further details, we refer the reader to~\cite{Etzion:13}.
\end{remark}
}

\subsection{Locality in the Subspace Metric}
\label{sec:locality-in-subspace}
In this section, we extend the concept of locality to that of {\it subspace-locality}. We begin with setting up the necessary notation. Let $\Omega\subseteq\Gq{M}{n}$ be a {Grassmannian code}. 
{To define the notion of subspace-locality, we need to to choose an ordered basis for every codeword subspace. It is possible to choose an arbitrary basis. However, 
we choose vectors in reduced column echelon form as an ordered basis since it turns out to be a natural choice for the lifting construction (described in Sec.~\ref{sec:lifting-construction}).}
Specifically, for every codeword $U\in\Omega$, consider an $M\times n$ matrix $\mymatrix{U}$ in a reduced column echelon form (RCEF) such that columns of $\mymatrix{U}$ span $U$. In other words, $\mymatrix{U} = \rcef{\mymatrix{U}}$ and $U = \subspace{\mymatrix{U}}$. Note that columns of $\mymatrix{U}$ form an ordered basis of $U$. 

For a set $S\subset[n]$, let $\project{\mymatrix{U}}{S}$ denote the $M\times |S|$ sub-matrix of $\mymatrix{U}$ consisting of the columns of $\mymatrix{U}$ indexed by $S$. Let $\project{U}{S} = \subspace{\project{\mymatrix{U}}{S}}$, and $\project{\Omega}{S} = \left\{\project{U}{S} : U\in\Omega\right\}$. Note that the code $\project{\Omega}{S}$ is essentially obtained by taking a projection of every subspace $U$ of $\Omega$ on the subspace formed by the basis vectors indexed by the elements in $S$. 

Now, we define the notion of subspace-locality in the following.
\begin{definition}[Subspace-Locality]
\label{def:local-subspace-code}
A {Grassmannian code} $\Omega\subseteq\Gq{M}{n}$ is said to have $(r,\delta)$ subspace-locality if, for each $i\in[n]$, there exists a set $\gam{i}\subset [n]$ such that
\begin{enumerate}
\item $i\in\gam{i}$,
\item $|\gam{i}| \leq r + \delta - 1$, 
\item $\dims{\project{\Omega}{\gam{i}}} = |\gam{i}|$, and
\item $\ds{\project{\Omega}{\gam{i}}} \geq \delta$.
\end{enumerate}
The code $\project{\Omega}{\gam{i}}$ is said to be the local code associated with the $i$-th basis vector for the subspaces of $\Omega$. A subspace code $\Omega\subseteq\Gq{M}{n}$ with minimum distance $d$ and $(r,\delta)$ locality is denoted as an $(M \times n, \log_{q}|\Omega|, d, r, \delta)$ {Grassmannian code}.
\end{definition}

\subsection{Grassmannian Codes with Subspace-Locality via Lifting} 
\label{sec:lifting-construction}
In~\cite{SilvaKK:08}, the authors presented a construction for a broad class of {Grassmannian codes} based on  rank-metric codes. The construction takes codewords of a rank-metric code and generates codewords of a {Grassmannian code} using an operation called {\it lifting}, described in the following. 

\begin{definition}[Lifting]
\label{def:lifting}
Consider the following mapping 
\begin{IEEEeqnarray}{rCcCl}
\Lambda & : & \GFmn{q} & \rightarrow & \Gq{m+n}{n},\nonumber\\
\label{eq:lifting}
{} & {} &  X & \mapsto & \Lambda(X) = \subspace{\twomatrix{I}{X}},
\end{IEEEeqnarray} 
where $I$ is the $n\times n$ identity matrix. The subspace $\Lambda(X)$ is called the {\it lifting} of the matrix $X$.\footnote{It is worth noting that the definition of the lifting operation is adapted to our notation. In~\cite{SilvaKK:08}, the authors define the lifting of an $m\times n$ matrix $X$ as the row space of the matrix $[I \:\: X]$, where $I$ is an $m\times m$ identity matrix. We define the lifting on columns, since rank-locality is defined with respect to columns.} Similarly, for a rank-metric code $\code \subseteq\GFmn{q}$, the subspace code $\Lambda(\code) = \left\{\Lambda(X) : X \in \code\right\}$ is called the {\it lifting} of $\code$. 
\end{definition}

Note that the lifting operation $X\mapsto\Lambda(X)$ is an injective mapping, since every subspace corresponds to a unique matrix in reduced column echelon form (RCEF). Thus, we have $|\Lambda(\code)| = |\code|$. Also, a subspace code constructed by lifting is a Grassmannian code, with each codeword having dimension $n$. 

The key feature of the lifting based construction is that the {Grassmannian code} constructed by lifting inherits the distance properties of its underlying rank-metric code. 
More specifically, we have the following result from~\cite{SilvaKK:08}. 

\begin{lemma}
\label{lem:lifting-distance}
(\cite{SilvaKK:08}) Consider a rank-metric code $\code\subseteq\GFmn{q}$. 
Then, we have
\begin{IEEEeqnarray}{rCl}
\ds{\Lambda(\code)} & = & 2\:\dr{\code}.\nonumber
\end{IEEEeqnarray} 
\end{lemma}

Next, we show that the lifting construction given in~\eqref{eq:lifting} preserves the locality property. 

\begin{lemma}
\label{lem:lifting-locality}
A {Grassmainnian code} obtained by lifting a rank-metric code with $(r, \delta)$ rank-locality has $(r,2\delta)$ subspace-locality.
\end{lemma}
\begin{IEEEproof}
Let $\code\subseteq\GFmn{q}$ be a rank-metric code with $(r, \delta)$ rank-locality. For each $i\in[n]$, there is a local code $\codeproj{\gam{i}}$ such that $\dr{\codeproj{\gam{i}}}\geq\delta$ due to the $(r,\delta)$ rank-locality of $\code$. 

Let $\Omega = \Lambda(\code)$ be the Grassmannian code obtained by lifting $\code$. Let $\project{\Omega}{\gam{i}} = \{\project{U}{\gam{i}} : U\in\Omega\}$. Consider a pair of codewords $V, V' \in\project{\Omega}{\gam{i}}$. Then, we have
\begin{equation*}
V = \subspace{\twomatrix{\hat{I}_{\gam{i}}}{C_{\gam{i}}}}, \quad V' = \subspace{\twomatrix{\hat{I}_{\gam{i}}}{C'_{\gam{i}}}},
\end{equation*}
where $\hat{I}_{\gam{i}}$ is an $n\times|\gam{i}|$ sub-matrix of the $n\times n$ identity matrix composed of the columns indexed by $\gam{i}$, and $C_{\gam{i}},C'_{\gam{i}}\in\codeproj{\gam{i}}$. Note that $\dims{V} = \dims{V'} = |\gam{i}|$. Thus, we have
\begin{IEEEeqnarray}{rCl}
\ds{V,V'} 
& \stackrel{(a)}{=} & 2\:\dims{V + V'} - \dims{V} - \dims{V'}\nonumber\\
& \stackrel{(b)}{=} & 2\:\dims{V + V'} - 2|\gam{i}|\nonumber\\
& \stackrel{(c)}{=} & 2\:\rnk{\begin{bmatrix} \hat{I}_{\gam{i}} & \hat{I}_{\gam{i}}\\ C_{\gam{i}} & C'_{\gam{i}} \end{bmatrix}} - 2|\gam{i}|\nonumber\\
& {=} & 2\:\rnk{\begin{bmatrix} \hat{I}_{\gam{i}} & 0\\ C_{\gam{i}} & C'_{\gam{i}} - C_{\gam{i}} \end{bmatrix}} - 2|\gam{i}|\nonumber\\
& = & 2\:\rnk{C'_{\gam{i}} - C_{\gam{i}}}\nonumber\\
\label{eq:ds-inequality}
& \stackrel{(d)}{\geq} & 2\delta,
\end{IEEEeqnarray}
where (a) follows from~\eqref{eq:subspace-distance} and the fact that $\dims{V+V'} = \dims{V} + \dims{V'} - \dims{V\cap V'}$, (b) follows due to $\dims{V} = \dims{V'} = |\gam{i}|$, (c) follows from the fact that for any pair of matrices $X$ and $Y$, we have 
$$\subspace{\left[{X} \:\:\:{Y}\right]} = \subspace{X} + \subspace{Y},$$
and (e) follows from $\dr{\codeproj{\gam{i}}}\geq\delta$. 

The result is immediate from~\eqref{eq:ds-inequality}.
\end{IEEEproof}
 
Now, by lifting rank-metric codes obtained via Construction~\ref{con:local-MRD-1}, we get a family of {Grassmannian codes} with locality. Specifically, from Lemmas~\ref{lem:lifting-distance} and~\ref{lem:lifting-locality}, we get the following result as a corollary.

\begin{corollary}
\label{cor:subspace-code-construction}
Let $\code_{Loc}$ be an $(m\times n, k, d, r, \delta)$ rank-metric code obtained by Construction~\ref{con:local-MRD-1}. The code $\Lambda(\code_{Loc})$ obtained by lifting $\code_{Loc}$ is an $((m+n) \times n, mk, 2d, r, 2\delta)$ {Grassmannian code}. 
\end{corollary}

{
\subsection{Application of Subspace-Locality in Networked Distributed Storage Systems}
\label{sec:networked-storage}
In this section, we present an application of Grassmannian codes with subspace-locality in distributed storage systems (DSS), in which storage servers are connected over a communication network that can introduce errors and erasures. We demonstrate how codes with subspace-locality can be helpful when users want to partially download the data stored on one or more racks, or when repairing a failed node. Fig.~\ref{fig:noisy-network} demonstrates an example for our set-up.

\begin{figure*}[!t]
\centering
\includegraphics[scale=0.6]{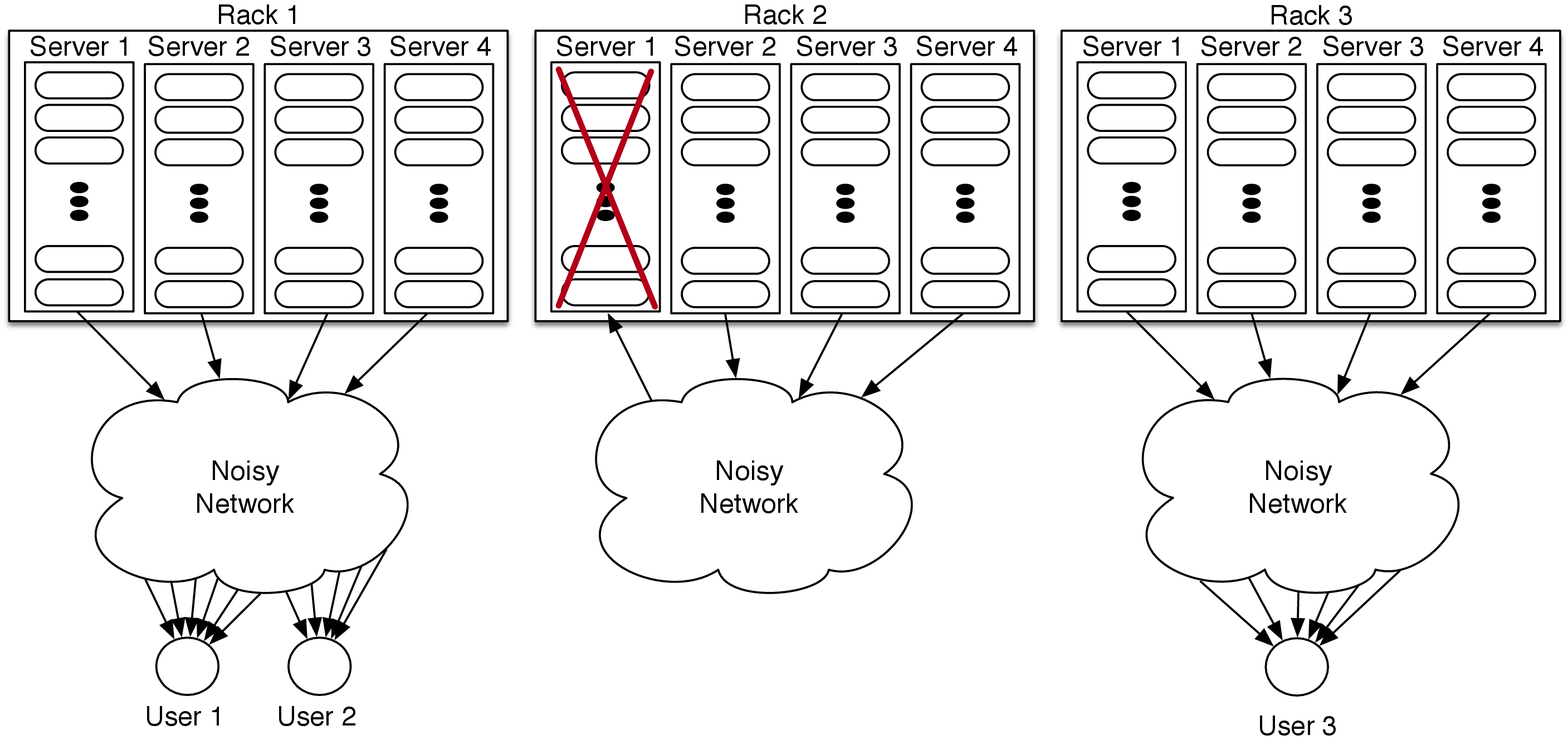}
\caption{We highlight a potential application of Grassmannian codes with subspace-locality in distributed storage systems, wherein storage servers can be accessed over a noisy network. 
In this example, we consider $n = 12$ servers located in $\mu = 3$ racks such that each rack contains $r+\delta-1 = 4$ servers.
Consider a scenario where users 1 and 2 are interested in downloading only the data stored on Rack 1. The nodes in the network use random linear network coding, and the network links can introduce errors and erasures. Subspace-locality ensures that the servers in Rack 1 can generate a Grassmannian code that is guaranteed to correct a certain number of errors and erasures introduced by the noisy network. Subspace-locality is also useful to repair a server when assessing other servers over a noisy network.
}
\label{fig:noisy-network}
\end{figure*}

For simplicity, we focus on the case of partial data download from a rack over a noisy network. Node repairs can be handled in a similar fashion. 
In particular, we consider the following set-up. Consider a DSS consisting of $n$ servers, which are located in $\mu$ racks such that each rack contains $r+\delta-1$ servers. Users can download data from the servers over a network that can introduce erasures and errors. 
Nodes in the network use random linear network coding to transfer {\it packets}~\cite{Ho:06}. Storage servers and users have no knowledge of the topology of the network or of the particular network code used in the network.\footnote{The goal of this section is to highlight the usefulness of subspace-locality for random linear network coding over a noisy network. A detailed study of various protocols for efficiently downloading data over a noisy network is beyond the scope of this paper.} 

We briefly mention the random linear network coding model, borrowing some notation from~\cite{SilvaKK:08}. 
Each link in the network can transport a {\it packet} of $M$ symbols in a finite field $\GF{q}$. Consider a node in the network with $a$ incoming links and $b$ outgoing links. The node produces an outgoing packet independently on each of its $b$ outgoing links as a random $\GF{q}$-linear combination of the $a$ incoming packets it has received.

Let us focus on a user $u$ interested in downloading the data stored on rack $j$, where $1\leq j\leq \mu$. We assume that the network contains $(r+\delta-1)$ mutually edge disjoint paths from the rack to the user. 

Suppose the data is encoded using an $((m+n)\times n, mk, 2d, r, 2\delta)$ Grassmannian code obtained using the lifting construction described in Sec.~\ref{sec:lifting-construction}. 
More specifically, first, the data is encoded using an $(m\times n, k, d, r, \delta)$ rank-metric code $\code$ as given in Construction~\ref{con:local-MRD-1}. Then, each of the $n$ servers stores a column of the codeword matrix. Let $C_{(j-1)(r+\delta-1)+i}$ denote the vector stored on the $i$-th server in the $j$-th rack. Let $I_l$ denote the $l$-th column of the $n\times n$ identity matrix. Then, each server $i$ in the $j$-th rack sends a packet $X^{(j)}_ i = \left[I_{(j-1)(r+\delta-1)+i}^T \:\: C_{(j-1)(r+\delta-1)+i}^T\right] \in \GF{q}^{1\times M}$ on its outgoing link, where $M = m+n$.

Let $X^{(j)}$ be an $(r+\delta-1) \times M$ matrix whose rows are the transmitted packets for rack $j$. We assume that the user collects $N$ $(\geq r)$ packets, denoted as $Y^{(u)}_1,\ldots,Y^{(u)}_N \in \GF{q}^{1\times M}$. Let $Y^{(u)}$ be an $N \times M$ matrix whose rows are the received packets. If the network is error free, then, regardless of the network topology, the transmitted packets $X^{(j)}$ and the received packets $Y^{(u)}$ can be related as $Y^{(u)} = AX^{(j)}$, where $A$ is an $N \times (r+\delta-1)$ matrix corresponding to the overall linear transformation applied by the network. 

Next, let us extend this model to incorporate packet errors and erasures. We consider that packet errors may occur at any link, which is a common assumption in the network coding literature. In particular, let us index the links in the network from $1$ to $\ell$. Let $Z_i$ denote the error packet injected at link $i \in \{1,\ldots, \ell\}$. If a particular link $i$ does not inject any error, then $Z_i$ is a zero vector. Let $Z$ be an $\ell \times M$ matrix whose rows are the error packets. Then, by linearity of the network code, we get
\begin{equation}
\label{eq:RLNC-channel}
Y^{(u)} = AX^{(j)} + BZ,
\end{equation}
where $B$ is an $N\times \ell$ matrix corresponding to the overall linear transformation applied by the network to the error packets. Note that the number of non-zero rows of $Z$ denotes the total number of error packets injected by the network. Further, the rank-deficiency of the matrix $A$ captures packet erasures caused by link failures.

Now, using~\cite[Theorem 1]{SilvaKK:08}, we immediately get the following result. 

\begin{proposition}
\label{prop:network-correction}
Suppose the network introduces at most $\rho$ erasures (\ie, the $\rnk{A} \geq r+\delta - 1 - \rho$), and injects at most $t$ error packets (\ie, the number of non-zero rows in $Z$ is at most $t$). Then, the user is guaranteed to recover the data from a rack provided
\begin{equation}
2t + \rho \leq \delta - 1.
\end{equation}
\end{proposition}
\begin{IEEEproof}
Let $\Omega_j = \Lambda(\code_j)$, where $\code_j$ is the $j$-th local code of $\code$. Note that $\subspace{[X^{(j)}]^T}\in\Omega_j$.  Further, from Corollary~\ref{cor:subspace-code-construction}, we have that $\ds{\Omega_j} = 2\delta$.

Now, the user can decode the data by using the minimum distance decoding rule as follows
\begin{equation}
\label{eq:min-subspace-distance-decoder}
\hat{X} = \arg\min_{X'\in\Omega_j} \ds{\subspace{X'},\subspace{[Y^{(u)}]^T}}.
\end{equation}
From~\cite[Theorem 1]{SilvaKK:08}, the decoding is guaranteed to be successful provided $2t + \rho < \ds{\Omega_j}/2$, from which the result follows.  
\end{IEEEproof}

\begin{remark}
\label{rem:lifting-for-networked-storage}
Note that, in general, Proposition~\ref{prop:network-correction} holds for any $(M\times n, \log_q|\Omega|, 2d, r, 2\delta)$ Grassmannian code $\Omega$ with disjoint local codes. 
In this case, during encoding, the first step is to fix an arbitrary injective mapping $\phi$ between data symbols and subspaces in $\Omega$. Then, given a set of data symbols to be stored, a subspace from $\Omega$ corresponding to the data symbols is obtained using the mapping $\phi$. Finally, each server stores a basis vector of this subspace.\footnote{Note that when a Grassmannian code is obtained via lifting, a server does not need to store the entire basis vector, but only the part due to the rank-metric code. This is because of the particular structure of the basis vectors obtained via lifting. On the other hand, for an arbitrary Grassmannian code, each server needs to store the entire basis vector. However, in typical applications, we have $m \gg n$, and the storage savings achieved by the lifting construction would be nominal.} During the partial data download, each server from the $j$-th rack transmits the stored basis vector as a {\it packet} on its outgoing link. 
\end{remark}

}

\section{Related Work and Comparison}

\subsubsection{\bf Codes with Locality} 
Codes with small locality were introduced in~\cite{Huang:07, Han:07} (see also~\cite{Oggier:11}). The study of the locality property was galvanized with the pioneering work of Gopalan \etal~\cite{Gopalan:12}, which established Singleton-like upper bound on the minimum distance of locally recoverable codes (LRCs). 
The distance bound has been generalized in multiple ways, see \eg,~\cite{Prakash:12,Papailiopoulos:14,Rawat:14,Kamath:14}. 
A large number of optimal code constructions have been presented, see \eg,~\cite{Silberstein:13,Tamo:LRC13,Ernvall:16,TamoB:14,Goparaju:14,Song:14,Huang:16}.  

Maximally recoverable codes (MRCs) are a class of LRCs that have the {\it strongest} erasure correction capability. 
The notion of maximal recoverability was first proposed by~\cite{Huang:07} and was generalized by~\cite{Gopalan:14}.

{LRCs as well as MRCs are primarily designed to correct small number of erasures locally. As an example, consider a family of distance-optimal LRCs presented in~\cite[Construction 8]{TamoB:14}.\footnote{We choose this construction because it requires the smallest possible field size (in particular $O(n)$) among the known constructions.} (See Sec.~\ref{rem:comparison-Tamo-Barg} for details.) 
Let $\code$ be an $(n,k)$ LRC from this family with $(r,\delta)$ locality. Let $\mu = n/(r+\delta-1)$, and $\code_1, \code_2, \ldots, \code_{\mu}$ denote the $\mu$ local codes with disjoint coordinates $C_1, C_2, \ldots, C_{\mu}$, respectively.
Then, a local code $\code_j$ can correct $\delta-1$ or less erasures in $C_j$ by accessing unerased symbols only from $C_j$ (for every $1\leq j\leq \mu$). Further, $\code$ can correct any $d-1$ erasures, where $d$ is the minimum distance given in the right hand side of~\eqref{eq:dist-bound-LRC}. An MRC can correct any erasure pattern that is information-theoretically correctable by any LRC with the same parameters.}

{Even though LRCs (and MRCs) are not designed to correct crisscross erasures, they can be easily adapted to correct crisscross erasure patterns. In particular, let us describe how an LRC can be adapted to mimic the performance of $\code_{Loc}$ given in Construction~\ref{con:local-MRD-1} for correcting crisscross erasures. Towards this, consider an $(mn,mk)$ LRC $\code^{LRC}$ with $(rm, (\delta-1)m+1)$ locality. Let $\mu = n/(r+\delta-1)$, and let $\code_1^{LRC}, \code_2^{LRC}, \ldots, \code_{\mu}^{LRC}$ denote the $\mu$ local codes with disjoint coordinates. Note that it is straightforward to construct such a code using ~\cite[Construction 8]{TamoB:14}.\footnote{Note that in this case the required field size would be $O(mn)$.} 

Suppose $mk$ data symbols are encoded using $\code^{LRC}$. 
The encoded symbols are arranged in an $m\times n$ array such that $(r+\delta-1)m$ symbols of $\code_j^{LRC}$ are arranged in columns $(j-1)(r+\delta-1)+1,\ldots,j(r+\delta-1)$, denoted as $C_j$. Note that $\code_j^{LRC}$ has the minimum Hamming distance $(\delta-1)m+1$. Therefore, $\code_j^{LRC}$ can locally correct any crisscross erasure pattern in $C_j$ of weight smaller than $\delta-1$. In fact, local codes $\code_j^{LRC}$ of $\code^{LRC}$ are stronger than the local codes $\code_j$ in $\code_{Loc}$. In particular, $\code_j^{LRC}$ can correct all erasure patterns in $C_j$ with fewer than $(\delta-1)m$ erasures, which include crisscross erasure patterns as a proper subset. 

On the other hand, despite their strong erasure correction capability, LRCs and MRCs are not capable of correcting crisscross and rank errors. This is because they are not guaranteed to have large rank distance.
}

\subsubsection{\bf Codes for Mixed Failures} 
Several families of codes have recently been proposed to encounter mixed failures. The two main families are: sector-disk (SD) codes and partial-MDS (PMDS) codes (see~\cite{Blaum:13, Plank:14, Blaum:14, Blaum:16}). Coded data are arranged 
in an $m \times n$ array, where a column of an array can be considered as a disk. Each row of the array contains $p$ {\it local} parities, and the array contains $h$ {\it global} parities. 
SD codes can tolerate erasure of any $p$ disks, plus erasure of any additional $h$ sectors in the array. PMDS codes can tolerate a broader class of erasures: any $p$ sector erasures per row, plus any additional $h$ sector erasures. {However, these codes cannot correct criscross erasures and errors.}

\subsubsection{\bf Codes for Correlated Failures}
Very recently, Gopalan \etal~\cite{Gopalan:17:SODA} presented a class of maximally recoverable codes (MRCs) for {\it grid-like topologies}. 
An MRC for a grid-like topology encodes data into an $m\times n$ array such that each row has $a$ {\it local} parities, each column has $b$ {\it local} parities, and the array has $h$ {\it global} parities. Such a code can locally correct any $a$ erasures in a row or $b$ erasures in column. When any $a$ rows and $b$ columns are erased, it can globally correct additional $h$ erasures. 
 
{MRCs for grid-like topologies can correct a large number of erasure patterns locally. However, their locality guarantees are significantly different. For instance, if an entire row (or less than $b$ rows) is erased, then it can be repaired by downloading $n-a$ symbols from any $m-b$ rows (similarly for column erasures). Further, these codes cannot correct crisscross and rank errors, as they are not guaranteed to have large rank distance.}

\subsubsection{\bf Rank-Metric Codes} 
Rank-metric codes were introduced by Delsarte~\cite{Delsarte:78} and were largely developed by Gabidulin~\cite{Gabidulin:85} (see also~\cite{Roth:91}). In addition, Gabidulin~\cite{Gabidulin:85} presented a construction for a class of MRD codes. Roth~\cite{Roth:91} introduced the notion of crisscross error pattern, and showed that MRD codes are powerful in correcting such error patterns. In~\cite{Blaum:00}, the authors presented a family of MDS array codes for correcting crisscross errors. {Existing constructions of rank-metric codes do not possess locality properties. In order to correct a criscross error/erasure pattern, it is required to read all the remaining symbols. To the best of our knowledge, this is the first work to propose the notion of locality in the rank metric.}

\subsubsection{\bf Subspace Codes} 
The important role of the subspace metric in correcting errors and erasures in non-coherent linear network codes was first noted in~\cite{KoetterK:08}. Since then, subspace codes (also known as codes over projective space) and constant-dimension subspace codes or Grassmannian codes have been studied in a number of research papers, see \eg,~\cite{Khaleghi:09,SilvaK:09,SilvaKK:08,EtzionS:09,GabidulinB:09,Gadouleau:10,EtzionV:11}, and references therein. {Existing constructions of Grassmannian codes do not possess locality properties. To the best of our knowledge, this is the first work to propose the notion of locality in the subspace metric.} 

\subsubsection{\bf Codes for Distributed Storage Based on Subspace Codes}  
Recently, subspace codes have been used to construct repair efficient codes for distributed storage systems. In~\cite{RavivE:15}, the authors construct regenerating codes based on subspace codes. In~\cite{SilbersteinES:17}, array codes with locality and availability (in the Hamming metric) are constructed using subspace codes. A key feature of these codes is their small locality for recovering a lost symbol as well as a lost column. {On the other hand, we present a construction of Grassmannian codes that have locality in the subspace metric. These codes are useful to recover partial data or repair nodes over noisy networks.}


\appendices

\section{Linearized Polynomials and Gabidulin Codes}
\label{app:lin-poly}
In this section, we first review some properties of linearized polynomials. (For details, please see~\cite{Lidl}.) Then, we specify Gabidulin codes construction. Let us begin with the definition of linearized polynomials. Recall that $x^{q^i} = \xq{x}{i}$.

\begin{definition}[Linearized Polynomial]
\label{def:lin-poly-2}
(\cite{Lidl}) A polynomial in $\GFm{q}[x]$ of the following form
\begin{equation}
\label{eq:lin-poly}
L(x) = \sum_{i=0}^{n} a_i \xq{x}{i}
\end{equation}
is called as a linearized polynomial or a $q$-polynomial over $\GFm{q}$. Further, $\max\{i\in[n] : a_i \neq 0\}$ is said to be the $q$-degree of $L(x)$ denoted as $\degq{L(x)}$.
\end{definition} 

The name arises from the following property of linearized polynomials, referred to as $\GF{q}$-linearity~\cite{Lidl}. Let $\mathbb{F}$ be an arbitrary extension field of $\GFm{q}$ and $L(x)$ be a linearized polynomial over $\GFm{q}$, then
\begin{IEEEeqnarray}{rCl}
\label{eq:F-q-linearity-1}
L(\alpha+\beta)  & = & L(\alpha) + L(\beta) \quad \forall \:\: \alpha, \beta\in\mathbb{F}.\\
\label{eq:F-q-linearity-2}
L(c\alpha) & = & cL(\alpha) \quad \forall \:\: c\in\GF{q} \:\: \textrm{and} \:\: \forall\:\:\alpha\in\mathbb{F}.
\end{IEEEeqnarray}

\begin{definition}[$q$-Associates]
\label{def:q-associates}
(\cite{Lidl}) The polynomials
\begin{equation}
\label{eq:q-associates}
l(x) = \sum_{i=0}^{n}c_i x^i \quad \textrm{and} \quad L(x) = \sum_{i=0}^{n}c_i \xq{x}{i}
\end{equation}
over $\GFm{q}$ are called $q$-associates of each other. In particular, $l(x)$ is referred to as the conventional $q$-associate of $L(x)$ and $L(x)$ is referred to as the linearized $q$-associate of $l(x)$.
\end{definition}

\begin{theorem}
\label{thm:lin-poly-roots}
\cite[Theorem 3.50]{Lidl} Let $L(x)$ be a non-zero linearized polynomial over $\GFm{q}$ and let $\GFext{q}{s}$ be the extension field of $\GFm{q}$ that contains all the roots of $L(x)$. Then, the roots form a linear subspace of $\GFext{q}{s}$, where $\GFext{q}{s}$ is regarded as the vector space over $\GF{q}$.
\end{theorem}

The above theorem yields the following corollary.

\begin{corollary}
\label{col:lin-poly-roots}
Let $L(x)$ be a non-zero linearized polynomial over $\GFm{q}$ with $\degq{L(x)} = l$, and let $\GFext{q}{t}$ be arbitrary extension field of $\GFm{q}$. Then, $L(x)$ has at most $l$ roots in $\GFext{q}{t}$ that are linearly independet over $\GF{q}$.
\end{corollary}

{\bf Gabidulin Code Construction:}
We review a class of maximum rank distance (MRD) codes presented by Gabidulin in~\cite{Gabidulin:85} for the case $m\geq n$. 
{Let $q$ be a prime power, let $m\geq n$, and let $\setP = \{\gamma_1, \cdots, \gamma_n\}\in\GFm{q}^n$ be $n$ linearly independent elements over $\GF{q}$. An $(n,k)$ Gabidulin code over the extension field $\GFext{q}{m}$ for $m\geq n$ is the set of evaluations of all $q$-polynomials of $q$-degree at most $k-1$ over $\setP$.} 

More specifically, 
let $G_{\m}(x)\in\GFext{q}{m}[x]$ denote the linearized polynomial of $q$-degree at most $k-1$ with coefficients $\m = [m_0 \:\: m_1 \:\: \cdots \:\: m_{k-1}]\in\GFm{q}^{k}$ as follows: 
\begin{equation}
\label{eq:Gabidulin-lin-poly}
G_{\m}(x) = \sum_{j=0}^{k-1}\mij{j}\xq{x}{j},
\end{equation}
Then, the Gabidulin code is obtained by the following evaluation map 
\begin{IEEEeqnarray}{rCcCl}
\label{eq:Gabidulin-eval-map}
Enc & :   & \GFext{q}{m}^k  & \rightarrow & \GFext{q}{m}^n\nonumber\\
{}     & {} & \m  & \mapsto & \left\{G_{\m}(\gamma), \gamma\in\setP\right\}
\end{IEEEeqnarray}
Therefore, we have 
\begin{equation}
\label{eq:Gabidulin-eval-map-2}
\code_{Gab} = \left\{\left(G_{\m}(\gamma),\gamma\in\setP\right) \mid \m\in\GFext{q}{m}^{k}\right\}.
\end{equation}

{\bf Reed-Solomon Code Construction:} It is worth mentioning the analogy between Reed-Solomon codes and Gabidulin codes. An $(n,k)$ Reed-Solomon code over the finite field $\GF{q}$ for $q\geq n$ is the set of evaluations of all polynomials of degree at most $k-1$ over $n$ distinct elements of $\GF{q}$. 
More specifically, let {$\setP = \{\gamma_1, \cdots, \gamma_n\}$} be a set of $n$ distinct elements of $\GF{q}$ ($q\geq n$). Consider polynomials $g_{\m}(x)\in\GF{q}[x]$ with coefficients $\m = [m_0\:\: m_1\:\: \cdots\:\: m_{k-1}]\in\GF{q}^{k}$ of the following form:
\begin{equation}
\label{eq:RS-poly}
g_{\m}(x) = \sum_{j=0}^{k-1}\mij{j}x^{j},
\end{equation}
Then, the Reed-Solomon code is obtained by the following evaluation map 
\begin{IEEEeqnarray}{rCl}
\label{eq:RS-eval-map}
Enc & : & \GF{q}^k \rightarrow \GF{q}^n\nonumber\\
& & \m \mapsto \left\{g_{\m}(\gamma), \gamma\in\setP\right\}
\end{IEEEeqnarray}
Therefore, we have 
\begin{equation}
\label{eq:RS-eval-map-2}
\code_{RS} = \left\{\left(g_{\m}(\gamma),\gamma\in\setP\right) \mid \m\in\GF{q}^{k}\right\}.
\end{equation}

\begin{remark}
\label{rem:comparison}
For the same information vector $\m = [m_0 \cdots m_{k-1}]\in\GF{q}^{k}$, the evaluation polynomials of the Gabidulin code and the Reed-Solomon code are $q$-associates of each other. 
\end{remark}

\section{Rank Distance Optimality}
\label{app:dist-optimality-proof}
We present a proof of the optimality of the proposed Construction~\ref{con:local-MRD-1} with respect to~\eqref{eq:dist-bound-rank-locality}. We use some properties of linearized polynomials which are listed in Appendix~\ref{app:lin-poly}.
We begin with a useful lemma regarding the minimum rank distance of a rank-metric code that is obtained through evaluations of a linearized polynomial.

\begin{lemma}
\label{lem:lin-poly-encoding}
Let $\setP$ be a set of $n$ elements in $\GFext{q}{m}$ that are linearly independent over $\GF{q}$ ($m\geq n$). Consider a linearized polynomial $L_{\m}(x)\in\GFext{q}{m}[x]$ of the following form
\begin{equation}
\label{eq:lin-poly-generic}
L_{\m}(x) = \sum_{j=1}^{k}\mij{i_j}\xq{x}{i_j},
\end{equation}
where $i_j$'s are non-negative integers such that $0\leq i_1 < i_2 < \cdots < i_k \leq n - 1$, and $k\leq n$. Consider the code obtained by the following evaluation map 
\begin{IEEEeqnarray}{rCl}
\label{eq:ev-map-generic}
Enc & : & \GFext{q}{m}^k \rightarrow \GFext{q}{m}^n\nonumber\\
& & \m \mapsto \left\{L_{\m}(\gamma), \gamma\in\setP\right\}
\end{IEEEeqnarray}
In other words, we have
\begin{equation}
\label{eq:eval-map}
\code = \left\{L_{\m}(\gamma) \mid \m\in\GFext{q}{m}^{k},\gamma\in\setP\right\}.
\end{equation}
Then, $\code$ is a linear $(m\times n, k, d)$ rank-metric code with rank distance $d \geq n - i_k$. 
\end{lemma}
\begin{IEEEproof}
First, note that a codeword $\cw\in\code$ is the evaluation of $L_{\m}(x)$ on $n$ points of $\setP$ for a fixed $\m\in\GFm{q}^k$. Thus, a codeword is a set of $n$ values each in $\GFm{q}$. By fixing a basis for $\GFm{q}$ as a vector space over $\GF{q}$, we can represent a codeword $\cw\in\GFm{q}^n$ as an $m\times n$ matrix $C\in\GF{q}^{m\times n}$. Thus, $\code$ can be considered as a matrix or array code.

Second, note that $\code$ is an evaluation map over $\GFm{q}$. Observe that $\m\mapsto L_{\m}(x)$ is an injective map. Since $q$-degree of $L_{\m}(x)$ is at most $n-1$, two distinct polynomials $L_{\m_j}(x)$ and $L_{\m_l}(x)$ result in distinct codewords, and thus, dimension of the code (over $\GFm{q}$) is $k$. 

Finally, we show that $\dr{\code} \geq n - i_k$. Notice that 
\begin{equation}
\label{eq:rank-dist-vs-deg}
\max_{L_{\m},\m\in\GFm{q}^k} \degq{L_{\m}} \leq i_k,
\end{equation}
where $\degq{F}$ denotes the $q$-degree of a linearized polynomial $F$. 

Consider a codeword $\cw$ as a length-$n$ vector over $\GFm{q}$. Let $\m_{\cw}$ be the message vector resulting in $\cw$, and $L_{\m_{\cw}}$ be the corresponding polynomial giving $\cw$. Let $C\in\GF{q}^{m\times n}$ be the matrix representation of $\cw$ for some basis of $\GFm{q}$ over $\GF{q}$. Suppose $\rnk{C} = w_r$. We want to prove that $w_r\geq n - i_k$. Suppose, for contradiction, $w_r < n - i_k$.

Let $\wt{\cw} = w$. Clearly, $w_r \leq w$. Without loss of generality (WLOG), assume that the last $n-w$ columns of $C$ are zero.  We know that $n-w$ points in $\setP$, $\left\{\gamma_{w+1},\ldots,\gamma_{n}\right\}$, are the roots of $L_{\m_{\cw}}(x)$. Note that, since elements of $\setP$ are linearly independent over $\GF{q}$, $w \geq n - i_k$ (see Corollary~\ref{col:lin-poly-roots} in Appendix~\ref{app:lin-poly}). 

WLOG, assume that the first $w_r$ columns of $C$ are linearly independent over $\GF{q}$. After doing column operations, we can make the middle $w - w_r$ columns as zero columns. Thus, there exist coefficients $c^l_j$'s in $\GF{q}$, not all zero, such that
\begin{equation}
\label{eq:roots}
\sum_{j=1}^{w_r}c_j^l L_{\m_{\cw}}(\gamma_j) + c_{w_r+1}^l L_{\m_{\cw}}(\gamma_{w_r+l}) = 0, \: \textrm{for} \: 1\leq l\leq w - w_r.
\end{equation}
By using $\GF{q}$-linearity property of linearized polynomials (see~\eqref{eq:F-q-linearity-1},~\eqref{eq:F-q-linearity-2}), the above set of equations~\eqref{eq:roots} is equivalent to
\begin{equation}
\label{eq:roots-2}
L_{\m_{\cw}}\left(\sum_{j=1}^{w_r}c^l_j \gamma_j + c^l_{w_r+1} \gamma_{w_r+l}\right) = 0, \quad \textrm{for} \:\: 1\leq l\leq w - w_r.
\end{equation}
Therefore, $\left\{\sum_{j=1}^{w_r}c^l_j \gamma_j + c^l_{w_r+1} \gamma_{w_r+l}, 1\leq l\leq w - w_r\right\}$ are also the roots of $L_{\m_{\cw}}(x)$. Together with $\left\{\gamma_{w+1},\ldots,\gamma_{n}\right\}$ as its roots, $L_{\m_{\cw}}(x)$ has $n - w_r > i_k$ roots. Note that, since $\gamma_j$'s are linearly independent over $\GF{q}$, so are all of the $n - w_r$ roots. Thus, $L_{\m_{\cw}}(x)$ has more than $i_k$ roots that are linearly independent over $\GF{q}$, which is a contradiction due to~\eqref{eq:rank-dist-vs-deg} and Corollary~\ref{col:lin-poly-roots}.
\end{IEEEproof}

From the above lemma, it follows that $\code$ obtained using Construction~\ref{con:local-MRD-1} is a linear $(m\times n,k)$ rank-metric code. Observe that the $q$-degree of $G_{\m}(x)$ is bounded as
\begin{IEEEeqnarray}{ll}
\degq{G_{\m}(x)}\nonumber\\ 
\: \leq \left(\frac{k}{r} - 1\right)(r + \delta - 1) + r - 1 = k - 1 + \left(\frac{k}{r} - 1\right)(\delta - 1).\nonumber
\end{IEEEeqnarray}
Hence, from Lemma~\ref{lem:lin-poly-encoding}, we have $\dr{\code} \geq n - k + 1 - \left(\frac{k}{r} - 1\right)(\delta - 1)$, which proves the rank distance optimality.

\end{document}